\documentclass[sigconf]{acmart}

\usepackage{amsmath,amsfonts}
\usepackage{hyperref}
\usepackage{amsmath} 
\usepackage{graphicx}
\usepackage{subfigure}
\usepackage{url}
\usepackage{color}
 \usepackage{algpseudocode}
\newcommand{\BfPara}[1]{{\noindent\bf#1.}\xspace}
\usepackage{makecell}
\usepackage{multirow}
\usepackage{float}
\usepackage{threeparttable}

\usepackage[inline]{enumitem}
\usepackage{booktabs}

\usepackage[T1]{fontenc}
\usepackage[utf8]{inputenc}
\usepackage{environ}
\usepackage{tikz}
\usetikzlibrary{arrows}
\usetikzlibrary{calc,matrix}
{\bfseries}{\itshape}
{\bfseries}{\itshape}
{\bfseries}{\itshape}
{\bfseries}{\itshape}
{\bfseries}{\itshape}
{\itshape}

\usepackage{epsfig}

\usepackage{xspace}
\usepackage{setspace}

\newcommand{\note}[1]{}

\newcommand{\etc}{{etc.}\xspace}
\newcommand{\eg}{{\em e.g.,}\xspace}
\newcommand{\ie}{{\em i.e.,}\xspace}

\newcommand{\etal}{{\em et al.}\xspace}

\usepackage{hyperref}
\definecolor{linkcolour}{rgb}{0,0.2,0.6}
\definecolor{xgreen}{rgb}{0.2,0.6,0.0}
\definecolor{xred}{rgb}{0.7,0.1,0.0}
\hypersetup{  
	pdfpagemode=pagewidth,
	plainpages=false, 
	colorlinks,  
	urlcolor=blue!50!black,   
	linkcolor=green!50!black,    
	citecolor=red!50!black,  
	bookmarksnumbered
}
\def\equationautorefname~#1\null{(#1)\null}

\colorlet{punct}{red!60!black}
\definecolor{background}{HTML}{ffffff }
\definecolor{delim}{RGB}{20,105,176}
\colorlet{numb}{magenta!60!black}

\usepackage{listings}
\setcopyright{none}
\renewcommand\footnotetextcopyrightpermission[1]{}
\definecolor{light-gray}{gray}{0.95}
\definecolor{darkgray}{rgb}{0.4, 0.4, 0.4}
\definecolor{editorGray}{rgb}{0.95, 0.95, 0.95}
\definecolor{editorOcher}{rgb}{1, 0.5, 0} 
\definecolor{editorGreen}{rgb}{0, 0.5, 0} 
\definecolor{orange}{rgb}{1,0.45,0.13}      
\definecolor{olive}{rgb}{0.17,0.59,0.20}
\definecolor{brown}{rgb}{0.69,0.31,0.31}
\definecolor{purple}{rgb}{0.38,0.18,0.81}
\definecolor{lightblue}{rgb}{0.1,0.57,0.7}
\definecolor{lightred}{rgb}{1,0.4,0.5}
\usepackage{upquote}
\usepackage{listings}

\definecolor{pblue}{rgb}{0.13,0.13,1}
\definecolor{pgreen}{rgb}{0,0.5,0}
\definecolor{pred}{rgb}{0.9,0,0}
\definecolor{pgrey}{rgb}{0.46,0.45,0.48}

\usepackage{balance}

\usepackage{tikz}

\begin{document}

\title{Understanding Internet of Things Malware by Analyzing Endpoints in their Static Artifacts}
 

\author{Afsah Anwar$^1$, Jinchun Choi$^{1,2}$, Abdulrahman Alabduljabbar$^1$, Hisham Alasmary$^{1,3}$, 
Jeffrey Spaulding$^4$, An Wang$^5$, Songqing Chen$^6$, DaeHun Nyang$^7$, Amro Awad$^8$, and David Mohaisen$^1$}
\affiliation{$^1${\em University of Central Florida}
\hspace{2mm}$^2${\em Texas A\&M University} \hspace{2mm}$^3${\em King Khalid University} \hspace{2mm}$^4${\em Canisius College}
\hspace{2mm}$^5${\em Case Western Reserve University}
\hspace{2mm}$^6${\em GMU}  \hspace{2mm}$^7${\em Ewha Womans University} \hspace{2mm}$^8${\em NCSU}}



\begin{abstract}
The lack of security measures among the Internet of Things (IoT) devices and their persistent online connection gives adversaries a prime opportunity to target them or even abuse them as intermediary targets in larger attacks such as distributed denial-of-service (DDoS) campaigns. In this paper, we analyze IoT malware and focus on the endpoints reachable on the public Internet, that play an essential part in the IoT malware ecosystem. Namely, we analyze endpoints acting as dropzones and their targets to gain insights into the underlying dynamics in this ecosystem, such as the affinity between the dropzones and their target IP addresses, and the different patterns among endpoints. Towards this goal, we reverse-engineer 2,423 IoT malware samples and extract strings from them to obtain IP addresses. We further gather information about these endpoints from public Internet-wide scanners, such as Shodan and Censys. For the masked IP addresses, we examine the Classless Inter-Domain Routing (CIDR) networks accumulating to more than 100 million ($\approx$78.2\% of total active public IPv4 addresses) endpoints.
Our investigation from four different perspectives provides profound insights into the role of endpoints in IoT malware attacks, which deepens our understanding of IoT malware ecosystems and can assist future defenses. 
\end{abstract}


\maketitle



\section{Introduction}\label{sec:introduction}
The Internet of Things (IoT) has reshaped the way in which people, businesses, and even cities interact with their environment through Internet-connected devices. There is no doubt that IoT devices have benefited the global economy and made our lives more efficient. With the number of IoT devices soaring into the tens of billions~\cite{statisticsIoT}, the potential adversaries have set their sights on these devices knowing that they are always connected. To this end, malicious code that targets IoT devices is on the rise that infects the device itself and receives code updates from dropzones around the world. Acting as intermediate nodes, these infected devices have the potential to launch attacks on other targets to form a massive distributed denial-of-service (DDoS) attack~\cite{MohaisenA14a, habibi2017heimdall, mahjabin2019load, frustaci2017evaluating}. 
Moreover, the majority of these IoT devices are at a high risk to the new threats due to the lack of security awareness among consumers and the lack of consensus on security standards among the IoT industry~\cite{Janakiram2016, shwartz2018reverse}.

Bastys \etal~\cite{BastysBS18} demonstrate that popular IoT app platforms are susceptible to attacks by malicious applet makers. With less than half of consumers changing default passwords on their IoT devices~\cite{Gemalto}, it is a no surprise that malware like Mirai has been able to amass a large botnet to launch massive DDoS attacks by simply using a dictionary of common IoT login credentials~\cite{AntonakakisABB17}. Compared to traditional hardware with operating systems with automated updates, IoT devices tend to have slower patch times and insecure communication~\cite{bertino2017botnets}. It makes them ``ideal targets'' for additional attacks like the \textbf{K}ey \textbf{R}einstallation \textbf{A}tta\textbf{ck} (KRACK) exploit~\cite{Ng2017}. It abuses design flaws in cryptographic Wi-Fi handshakes to reinstall existing keys which allows attackers to eavesdrop on network traffic or even inject malicious content~\cite{KRACK_vanhoef-ccs2017}. 

Alrawi \etal~\cite{AlrawiLAM19,alrawi21circle} revisited the literature and evaluated security of IoT devices and software, unveiling various outstanding in the existing ecosystem that could be resolved with existing solutions. With the proliferation of IoT devices in today's world, we even see decades-old attacks resurface to take advantage of vulnerable IoT devices~\cite{alrawi21circle}. For example, the SSHowDowN Proxy attack discovered by Akamai~\cite{SSHowDowN_Akamai16} utilizes a 12-year old vulnerability in OpenSSH to effectively take over the device to remotely generate attack traffic. 

The impact of IoT malware is significant. Compromising IoT devices at scale, adversaries form a network of botnets to launch major attacks. For example, recently, 13,000 compromised IoT devices were used to generate persistent traffic of 30 Gbps, targeting numerous financial institutions, with significantly low intensity than the recorded Mirai botnet attack that generated devastating attack traffic of 620 Gbps~\cite{FinBusinessDDoSAttack}. Similarly, a service provider in the US survived the largest DDoS attack with attack traffic staggering to 1.7 Tbps~\cite{USlargestDDoS}.

Realizing the susceptibility of the IoT devices, malware authors can exploit such weaknesses to employ them as intermediary targets for large attacks~\cite{XuWP14,TrendMicro}. For an attack to be successful, the Command and Control (C2) servers, the intermediary targets, and the victim must be connected to the Internet, thereby making it essential to study these endpoints.

In this work, we extract endpoints from IoT malware samples by reverse-engineering those samples and perform a data-driven study to analyze their different traces, such as geographical affinities, organizations, ports, and their exposure to attacks. Taking our earlier work~\cite{ChoiAAS19} forward, this work dives deeper into exploring the geographical affinity between endpoints by studying their region- and city-level characteristics. Additionally, this work explores the devices characteristics (behind endpoints) accessible through the public Internet that are covered by the attack purview. Our work is essential to understand the Indicators of Compromise (IoCs) and the behavioral aspects of the targets that can be used for threat intelligence or threat hunting. For the masked IP addresses in the malware, we analyze the Classless Inter-Domain Routing (CIDR) addresses that accumulate to more than 100.7 million IP addresses accumulation to $\approx$78.2\% of total IPv4 addresses. We calculate the ratio with respect to the total responsive public IPv4 addresses as observed using Censys~\cite{DurumericAMB15}.

\begin{figure}[t!]
	\centering
	\includegraphics[width=0.95\linewidth]{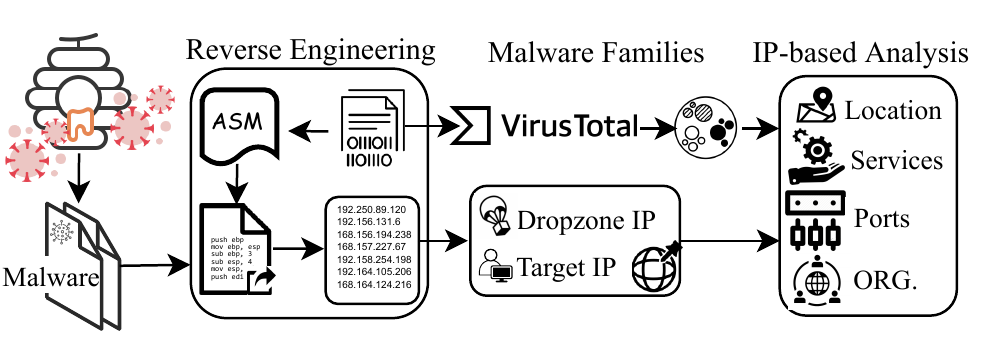}\vspace{-1mm}
	\caption{Data collection and extraction system. We start by collecting malware binaries from IoTPOT~\cite{Pa2016_IoTPOT}. The dataset is reverse engineered to extract the embedded IP addresses, which is then categorized into dropzones and target IP addresses. Using VirusTotal and Geolocation APIs, we extract additional behavioral information of the IP addresses. ORG.: organization information of the collected IP addresses.}
	\label{fig:system}
\end{figure}

\BfPara{Contributions} The main goal of this study is to analyze the dynamics exposed by the affinities among endpoints. A successful attack environment encompasses endpoints that act as the enablers. They help inflate the attack impact by acting as sources for the expansion of the network botnets. Practically, these endpoints are used to obtain instruction for infection, such as downloading shell scripts, malware binaries, and launching a flood attack~\cite{alrawi21circle}. The creation of an attack environment also requires endpoints that are targeted with an intention to create or expand a network of infected devices. Such devices can be selected algorithmically (using domain generation algorithms) or by use of a static and exhaustible list~\cite{ThomasM14}. Emphasizing on the enabling and the targeted endpoints, we make the following contributions. 
\begin{enumerate} 
    \item We analyze the dropzone-target inter-relationships. Specifically, we investigate the target IP addresses among different dropzone IP addresses.
    \item We perform a geographical analysis of the dropzones and targets. Towards this, we analyze the locations of dropzones and their target IP addresses.
    \item We perform a network penetration analysis of the targets and dropzone IP addresses. Specifically, we analyze risks associated with the IP addresses through insights gained from Shodan, an Internet-connected devices search engine.
    \item We analyze the attack exposure of networks and IP addresses. For masked target endpoints, we examine the entire network and study the network devices and their exposure to risk.
\end{enumerate}
The results of our analysis provide insightful findings about the role of endpoints in IoT malware ecosystems, which can also assist effective defenses in the future.

\BfPara{Organization} We describe our dataset, data augmentation and pipeline, and outline our goals and objectives in section~\ref{sec:dataGoal}. In section~\ref{sec:ipCentric}, we perform IP address-centric analysis of the endpoints followed by a network-centric analysis in section~\ref{sec:networkCentric}. We review the related work in section~\ref{sec:related} and discuss the implications of the results in section~\ref{sec:discussion}. Finally, we present conclusions and future work in section~\ref{sec:conclusion}.

\section{Dataset and Goals}\label{sec:dataGoal}

We describe our dataset and its augmentation, and describe the goals and objectives of this work. 
We reverse-engineer the malware and extract endpoints from the strings. We then add information related to the endpoints from different sources to facilitate our analyses. Fig.~\ref{fig:system} presents the process at a high-level; we explain the process in the rest of the section.

\subsection{Dataset}\label{dataset}
We were faced with the difficulty of obtaining IoT malware samples and turned to one of the first honeypots specifically targeted towards IoT threats. Proposed in 2015 by Pa \etal~\cite{Pa2016_IoTPOT}, the IoTPOT honeypot emulates the Telnet services (later improved to include other services) of different IoT devices and communicates with a back-end component called IoTBOX, which operates multiple virtual environments including eight different CPU architectures (\eg MIPS, ARM, \etc). We obtained a total of 2,423 IoT malware samples, which were graciously given to us by the authors of IoTPOT. The dataset represents four different malware families, labelled by augmenting the results from VirusTotal (VT) and by using AVClass~\cite{AVClass}. For malware samples that do not have a decisive family label from the VT results, those malware samples are labeled as SINGLETON.  The distribution of malware families can be seen in Table~\ref{tab:mal_family}.

\begin{table}[t]
\caption{Distribution of malware by family. DZ stands for Dropzone.}
\label{tab:mal_family}
\setlength{\tabcolsep}{0.35em}
\centering
\begin{tabular}{|l|c|c!{\vline width 2pt}l|c|c|}
\Xhline{2\arrayrulewidth}
Target family & Count & \% & DZ family & Count & \% \\ \Xhline{2\arrayrulewidth}
Gafgyt        & 930 & 95.58   & Gafgyt          & 2,294 & 98.96 \\ \hline
Tsunami       & 39 & 4.01   & Tsunami         & 24  & 1.04  \\ \hline
SINGLETON     & 3  & 0.31    & -                & -   & -   \\ \hline
Lightaidra    & 1  & 0.10   & -                & -   & -   \\
\Xhline{2\arrayrulewidth}
\end{tabular}
\end{table}

We reverse-engineer and analyze the malware samples using Radare2~\cite{radare2}, an open-source malware analysis framework. We find strings in the malware binaries, especially IP addresses, and classify those addresses by their association with special keywords into two classes: {\tt dropzone} and {\tt target} IP addresses, defined as follows:

\begin{itemize}[leftmargin=2.5mm,itemsep=3pt,topsep=3pt,parsep=0pt,partopsep=0pt]
    \item \BfPara{Dropzone IP} Adversaries often keep malware binaries in remote servers to distribute them after gaining access to victim devices. These remote servers are identified by {\tt dropzone} IP addresses, controlled and managed by adversaries and used for malware propagation and management. As such, the dropzone IP addresses are associated with wget, HTTP, TFTP, GET, or FTP in the residual strings obtained from the malware analysis.
    \item \BfPara{Target IP} To infect victim hosts, the malware uses a list of IP addresses, including target devices. We refer to these IP addresses as {\tt targets}. We note that a large number of those target addresses in our analysis are masked. For example, 123.17.*.* is one of the target IP address that is masked at /16; the attacker can utilize this address targeting all IPs in the network address space.
\end{itemize}

\begin{figure}[t!]
	\centering
	\includegraphics[width=0.75\linewidth]{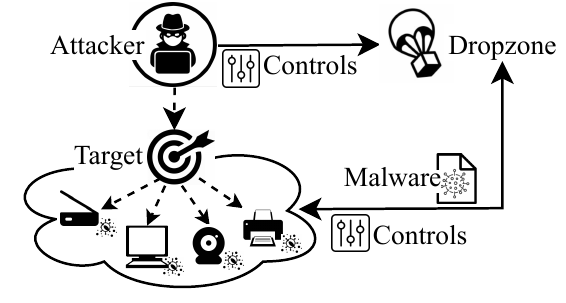}\vspace{-2mm}
	\caption{An overview of the dropzone and target in the malware life-cycle. The attacker uses a pre-configured or infected server (dropzone) for attack propagation. The attacker tries to propagate to targets by exploiting weaknesses. The malware accesses the dropzone and downloads scripts and payload to the target. The infected target devices repeat the process.}
	\label{fig:overview} 
\end{figure}

We find the internal network addresses (\eg 192.168.*.*), loopback addresses (\eg 127.0.0.1) from our target dataset and remove them, since they are irrelevant to our analysis. Also, we note that the Mirai source code contained a list of ``don't scan'' addresses, including various U.S. Department of Defense (DoD) address blocks, as well as internal addresses~\cite{impervaMirai}, which we exclude.  
Fig.~\ref{fig:overview} shows the dropzone and target in the malware life-cycle, including dropzone setting, victim host (target) compromise, and download of malware from the dropzone to the target.

\BfPara{Data Augmentation} 
We group the target and dropzone addresses by malware. Since a dropzone can be used by multiple malware, and to analyze the overall sample-space a dropzone caters to, we cluster the target IPs by each dropzone.

Using \textit{UltraTools}~\cite{ultratools}, a free Domain Name Server (DNS) and domain lookup tool, and \textit{Censys}~\cite{censys}, each of the targets and dropzones is augmented with country, ASN (Autonomous System Number), and location (\eg latitude and longitude). Using Shodan~\cite{shodan} we also obtain vulnerable endpoints on the Internet that are susceptible to targeted attacks. We correlate the results obtained from Shodan with our target and dropzone addresses to augment those addresses with additional information, such as vulnerabilities in services on those addresses, operating systems used running on top of them, and open ports (\ie services running on the addresses).

We observe some dropzone addresses have no current information, \eg they are no longer connected to the Internet. This confirms that the dropzones are short-lived---long enough to carry out an attack and short not to be detected. As such, we leverage historical data of those IP addresses from Shodan to determine the necessary data points associated with them.

\BfPara{Data Overview}
The distribution of the target addresses exposes family-level affinities by highlighting what set of addresses is being targeted by different malware. In particular, we observe a total of 106,428 target IP addresses, resulting in 2,211 \textit{unique target} IP addresses associated with 973 malware samples. This makes the analysis of affinities, by understanding what makes these target IP addresses the favorite among malware authors, of paramount importance.

The use of dropzone IP addresses by the malware exhibits that the malware shares dropzones among themselves, with some contacting multiple dropzones for commands. In particular, for the dropzones, we find that 877 \textit{unique dropzone} IP addresses are being shared by 2,318 malware samples with 2,407 occurrences. Moreover, while we successfully extract dropzone IP addresses from the majority of malware samples, we can find target IP addresses only in fewer malware samples. This shows a thought pattern of malware authors, \ie while they share the dropzone IP addresses in the static code, they do not reveal the target IP addresses. This can be because they obfuscate this part of the code, employ domain generation algorithms, or use a custom list of IP addresses in the downloaded binary file (\ie DNS.txt) from a dropzone at runtime, as shown in Fig.~\ref{fig:target_list}.

We also notice that 40\% of the malware samples contain target IP addresses, while 95.66\% of them contain dropzones in their strings. We also observe disassembled codes of malware samples that have dropzone but no targets, which is explained either: (i) code-based generation of IP addresses, rather than static IP address listing~\cite{AndersonWF16}, and (ii) packing.

\begin{figure}[t]
    \centering
\begin{lstlisting}[backgroundcolor = \color{blue!75!red!10!gray!30}, framexleftmargin = 1em,  aboveskip=0pt,
  belowskip=0pt, 
  frame=t, framesep=0.15cm, framerule=0pt]
wget \%s -q -O DNS.txt 
    || busybox wget \%s -O DNS.txt 
    || /bin/busybox wget \%s -O DNS.txt 
    || /usr/busybox wget \%s -O DNS.txt
\end{lstlisting}
    \caption{Retrieving a list of target hosts.}
    \label{fig:target_list}
\end{figure}

\subsection{Goals and Objectives}\label{goals}
The objective of this work is to conduct a comprehensive analysis of endpoints in IoT malware, including sources, C2's, intermediary targets, and victims. In particular, we formulate questions that we answer through a data-driven analysis to find the correlations between the endpoints of the dropzones and targets, addressing the following goals:
\begin{itemize}[leftmargin=2.5mm,itemsep=3pt,topsep=3pt,parsep=0pt,partopsep=0pt]
\item \BfPara{Dropzone-Target Inter-Relationships}
Since malware associated with certain dropzones point to specific target IP addresses, could these IP addresses be similar or identical to the addresses of targets in other dropzones?
To answer this, we reverse-engineered and analyzed the malware's disassembly to extract all target IP addresses for each dropzone.
\item \BfPara{Geographical Analysis}
What are the characteristics of the areas where the dropzones are located? How does this affect the distribution of dropzones and targets? 
For that, we analyze the distribution of the distance between the dropzones and their targets, and examine the distribution from various perspectives at the country, state, and city level.
\item \BfPara{Network Penetration Analysis}
What are the vulnerable services used for both dropzones and target IP addresses? Which organizations own these addresses?
We analyze attributes of dropzone and target IP addresses such as their active network ports, organization, and known vulnerabilities from Internet-wide scanners; Shodan and Censys~\cite{censys}.
\item \BfPara{Attack Exposure}
How exposed are the IP addresses in the target's network address space? 
Towards this, we analyze the targets and look for vulnerabilities in the services that they use. For the masked targets, we analyze the network space and examine their up-to-date susceptibility.
\end{itemize}

To answer these questions, we divide our data-driven analysis into (i) IP centric analysis and (ii) network centric analysis. We cover those directions in the two following sections.

\begin{figure}[t!]
	\centering
	\includegraphics[width=0.85\linewidth]{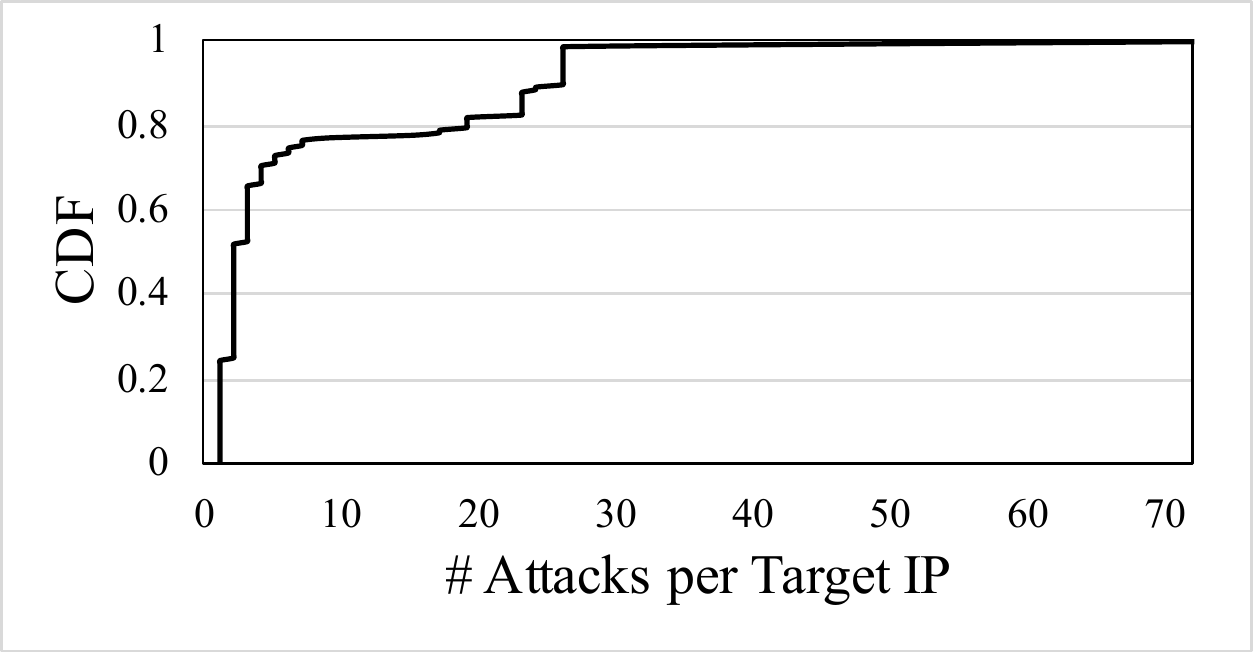}
	\vspace{-2mm}
	\caption{CDF graph of the number of attacks in the unique target IP. Most of them are targeted less than 10 times.}\
	\label{fig:CDF_attacked_target_IP} \vspace{-2mm}
\end{figure}

\section{IP Centric Analysis}\label{sec:ipCentric}

In this section, we analyze the dropzone-target inter-relationship, conduct a country-, region-, and city-level geographical analysis of IP addresses, and perform a penetration analysis of the IP addresses to examine their susceptibility.

\subsection{Dropzone-Target Inter-relationship}
To inspect the dropzone-target relationship, we examine the affinity between the dropzone and the target IP addresses.
Fig.~\ref{fig:CDF_attacked_target_IP} shows the cumulative distribution function (CDF) of the number of attacks that a unique target IP address receives.
While $\approx$77\% of the unique target IPs received less than 10 attacks, one unique target IP received 72 attacks. 

Throughout this research, we found one dropzone IP (\texttt{50.115.166.193}) that was only associated with one malware sample, and that sample pointed to 1,265 network addresses, which was significantly larger than the average of 121 target IP addresses for a typical malware sample. Also, they are masked network addresses (most of them are /16 masked, as mentioned in Table~\ref{tab:tg_composition}), which means that one target network address can be larger dynamically. Conversely, the dropzone IP (\texttt{5.189.171.210}) has 86 associated malware samples, but each of those points to a single target IP address.

\begin{figure}[t]
	\centering
	\includegraphics[width=1.0\linewidth]{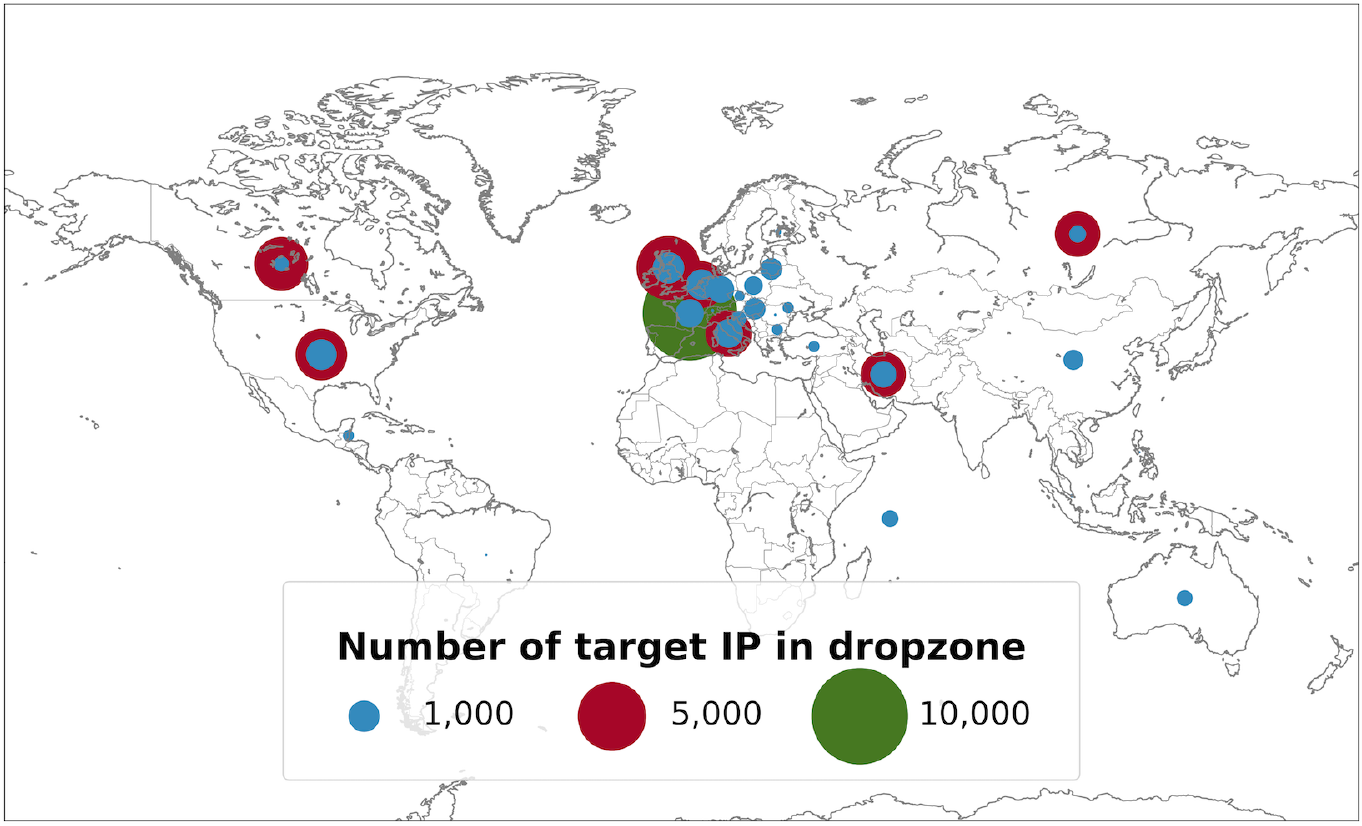}
	\caption{Distribution of dropzones by the number of target IPs. The location of each circle is based on the country extracted from the dropzone IP lookup information.}
	\label{fig:Distribution_dz} 
\end{figure}

\begin{table}[t]
\caption{Top 5 dropzone IP{\upshape s} per the number of targets.}
\label{tab:dz_number_target}
\small
\centering
\begin{tabular}{|c!{\vline width 1pt}l|c|c|}
\Xhline{2\arrayrulewidth}
Rank & Dropzone IP & Malware & Total Targets \\ \Xhline{2\arrayrulewidth}
1 & 163.172.104.150 & 35                 & 9,529                    \\ \hline
2 & 145.239.72.250  & 22                 & 5,632                    \\ \hline
3 & 45.76.131.35    & 17                 & 4,352                    \\ \hline
4 & 64.137.253.50   & 26                 & 3,066                    \\ \hline
5 & 198.175.126.89  & 11                 & 2,816                    \\ 
\Xhline{2\arrayrulewidth}
\end{tabular}
\end{table}

\begin{figure}[t]
\begin{center}
    \subfigure[CDF graph of the number of overlapped target IPs.]{\includegraphics[width=0.45\linewidth]{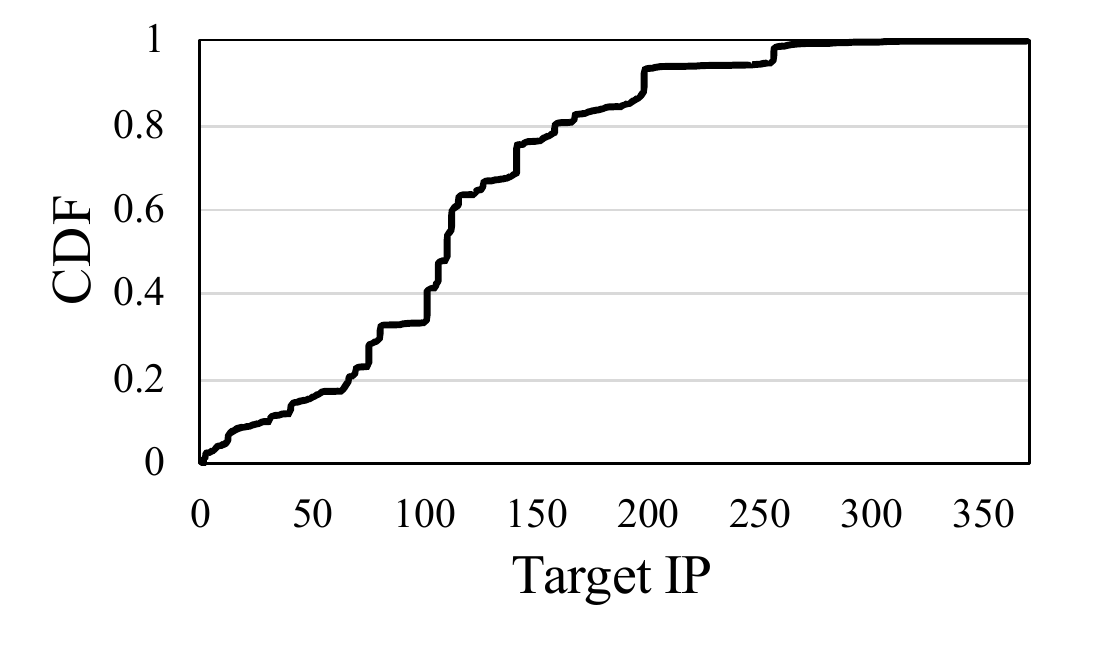}\label{fig:CDF_relation_overlap_IP}}
    \subfigure[The ratio of the number of overlapped target IPs to the number of unique target IPs in dropzones.]{\includegraphics[width=0.47\linewidth]{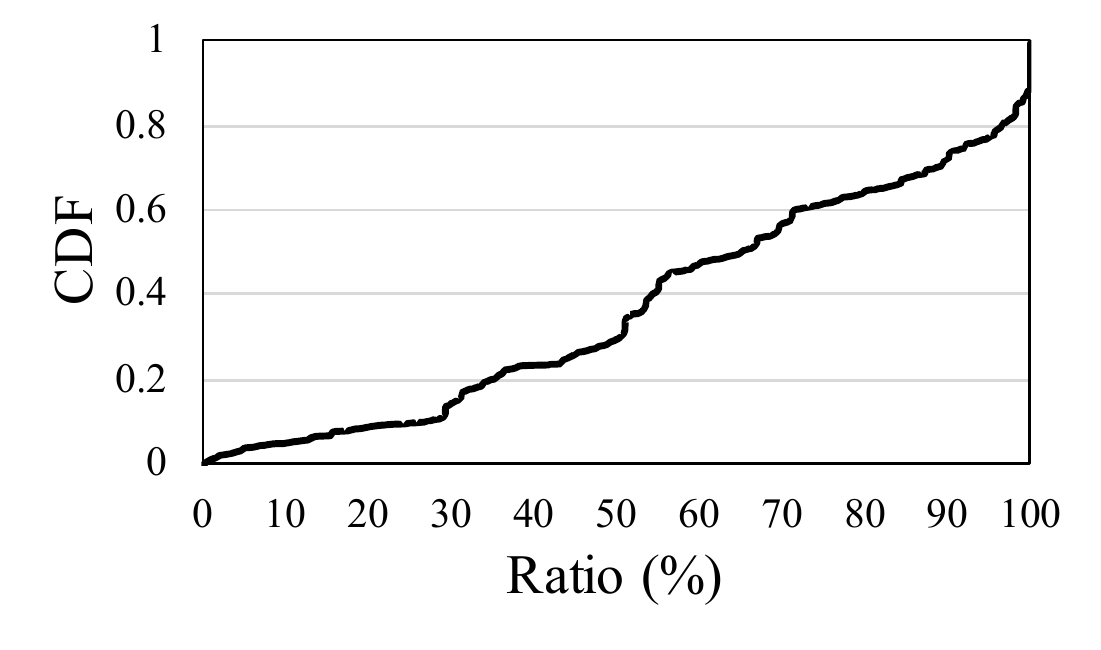}\label{fig:CDF_relation_overlap_ratio}}
    \caption{CDF graphs showing the distribution of the number of overlapped target IP addresses and their ratio.}
    \label{fig:CDF_relation_overlap}
\end{center}
\end{figure}

\BfPara{Geographical Distribution} Fig.~\ref{fig:Distribution_dz} plots the location of dropzone IP addresses around the world and the distribution by the number of the target IP addresses they are associated with. As shown in Fig.~\ref{fig:Distribution_dz}, dropzones can be found distributed mainly in North America and Europe. 
Moreover, through our further analysis we found that the first IP address (\texttt{163.172.104.150}) (Table~\ref{tab:dz_number_target}) is associated with 35 malware samples affecting 9,529 target IP addresses.

\BfPara{Shared Targets Between Dropzones}
To inspect the shared targets between dropzone IP addresses, we group the dropzone IP addresses and report the common (overlapping) targets among the dropzones. Since dropzones can be associated with multiple instances of malware, each malware can have its own list of target IP addresses. If we assume that a dropzone has a union of target IPs for each malware belonging to that particular dropzone, we can aggregate all of their target IPs into a larger set of target IPs. We denote $U_{dz}$ as the union of all target IPs for a particular dropzone.
To analyze the overlapping target IP addresses and understand the criteria for choosing target addresses, we compare $U_{dz}$ of each combination of dropzone addresses from a dataset of 877 unique dropzone addresses. Using combinations, we found combinations of $$\dbinom{877}{2} = \frac{877!}{2!(877-2)!} = 384,126.$$ Upon removing 365,968 cases that do not have common target IP between them, we reduce the combination to 18,158 dropzone IP pairs. This dataset of 18,158 dropzone IP pairs is a combination of only 247 unique dropzone IP addresses from the dataset of 877 unique dropzone IP addresses.

We found 71 cases that had more than 300 overlapped target IPs in Fig.~\ref{fig:CDF_relation_overlap_IP}. Fig.~\ref{fig:CDF_relation_overlap_ratio} shows that there were 2,199 cases (12.11\%) which are 100\% overlapped between dropzones. Overall, we found 6,451 cases (35.53\%) in which the overlap was more than 80\%.
In Table~\ref{tab:overlap_relation_overlap_frequency}, we list the top 5 cases of the number of overlapped target IP addresses by the number of dropzone pairs and the percentage that these pairs are from the total of 18,158 dropzone IP pairs. Notice that these top 5 cases of dropzone pairs make up over 30\% of all cases.

\BfPara{Summary} It is evident from the results of the above analysis that a large number of targets are being shared between dropzones. If the target IP addresses between different dropzones are matched 100\%, it is possible that the attacker obtained the same targets through similar vulnerability analysis (\ie Shodan) or shared the target list from other attackers through underground communities. For example, a simple search of ``default password'' on Shodan gives 69,093 results. Additionally, a sizable match suggests that the target list may have been partially shared, or the attacker may have added or removed certain targets to the list for directed attacks.

\begin{table}[t]
\caption{Top 5 Cases of the number of overlapped targets by the number of dropzone pairs.}
\label{tab:overlap_relation_overlap_frequency}
\setlength{\tabcolsep}{0.6em}
\centering
\begin{tabular}{|c!{\vline width 1pt}c|c|c|}
\Xhline{2\arrayrulewidth}
Rank & Overlap Targets & Dropzone Pairs & \% \\ \Xhline{2\arrayrulewidth}
1 & 101     & 1,385  & 7.63 \\ \hline
2 & 141     & 1,230  & 6.77 \\ \hline
3 & 110     & 1,078  & 5.94 \\ \hline
4 & 106     & 1,028  & 5.66 \\ \hline
5 & 75      & 1,000  & 5.51 \\ 
\Xhline{2\arrayrulewidth}
\end{tabular}
\end{table}

\subsection{Geographical Analysis}
\begin{figure}[t]
	\centering
	\includegraphics[width=.8\linewidth]{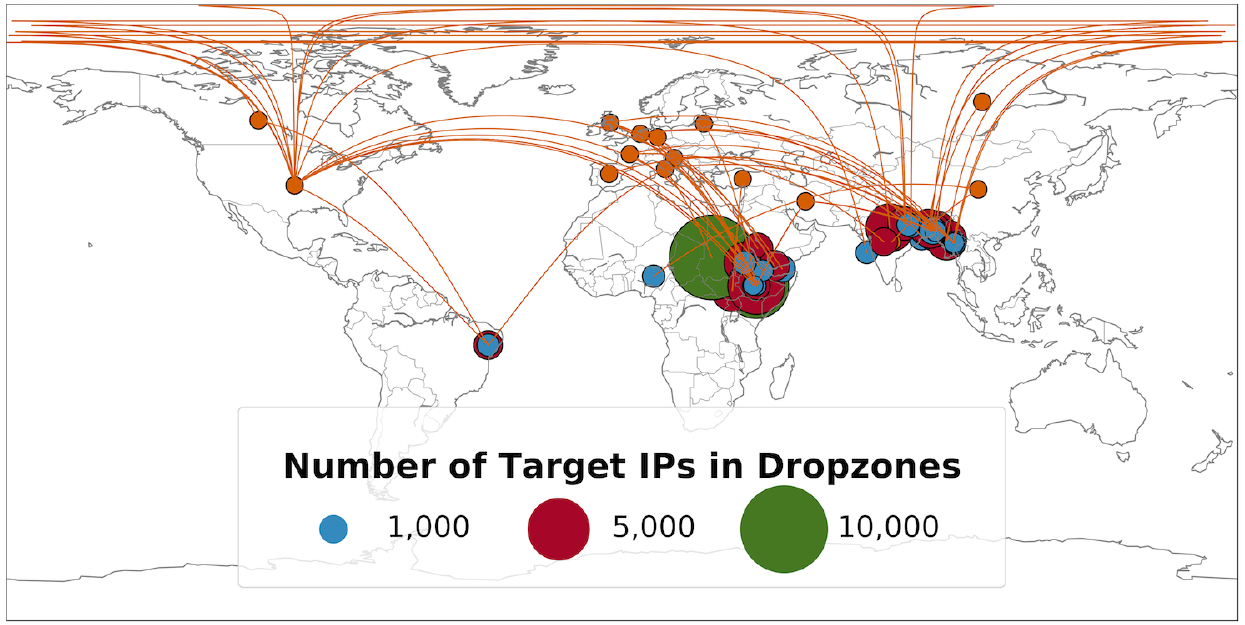}
	\caption{Attacks trends between dropzones and target IPs. We only plot attacks that have over 500 target IPs. The orange circle represents dropzones, and blue, red, and green circles stand for target areas.}
	\label{fig:trend_attack}
\end{figure}

In this section, we focus on the distribution of the distances between the dropzones and their target IPs. It will be apparent that a large number of dropzone-target pairs have a certain range of distances, which is related to the distribution of dropzones and their targets in each country. For example, we noticed that there are several target IPs located in Vietnam, Brazil, and China. 
To visualize the flow of attacks in a holistic sense, we plotted circular areas whose sizes are proportional to the number of targets and are placed according to their location on a world map with geodesic lines originating from various dropzone locations (see Fig.~\ref{fig:trend_attack}).

\begin{figure}[t]
	\centering
	\includegraphics[width=0.8\linewidth]{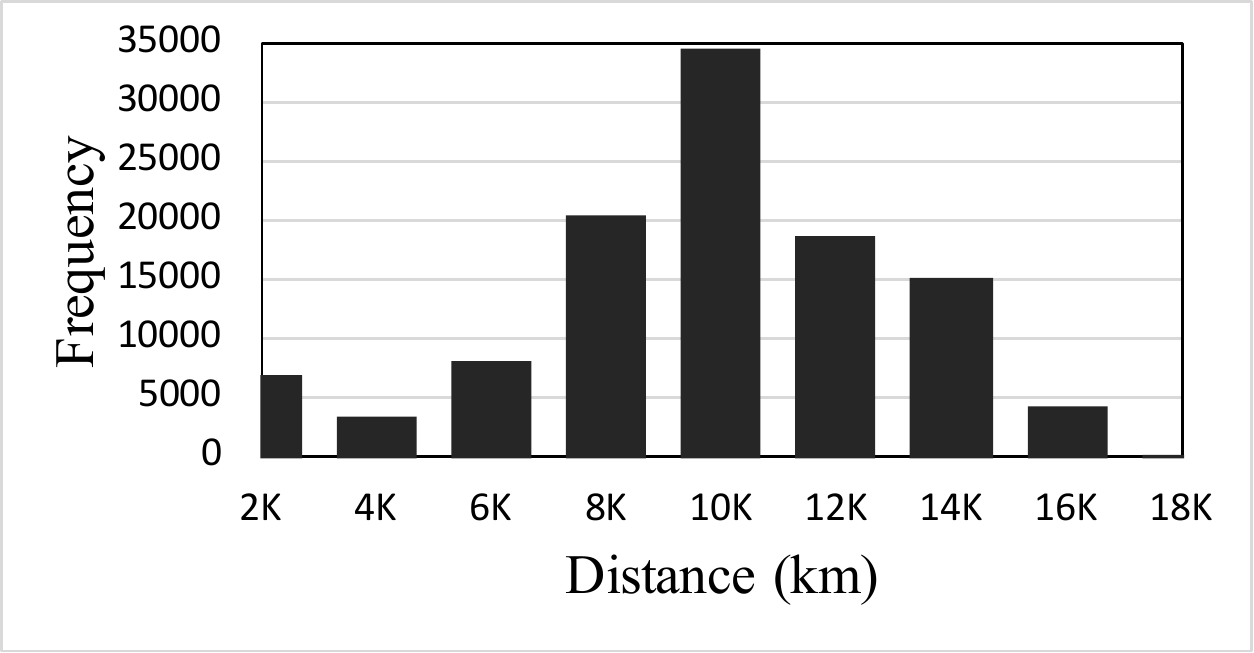}
	\caption{Histogram graph of distances between dropzone and target IPs. One bar represents the number of dropzone-target pairs with a distance that has a value within the range of the x-axis.}
	\label{fig:Hist_distance}
\end{figure}

\BfPara{Distance Between Dropzone and Target}
As mentioned previously, a dropzone IP can be associated with several malware instances where each malware can point to one or more target IPs. Knowing the locations of these IPs, we calculate the distance between the dropzone and its target if they are related to the same malware instance. Each distance shows the locality of the attack. The total number of calculated distance cases is 111,480.
Fig.~\ref{fig:Hist_distance} presents an alternate view with a histogram plot of the distances between the dropzones and their target IPs.

Our result of the majority of the distance shows the 8K-10K km range had the most frequent number of cases totaling 34,479 (30.93\% of all dropzone-target distance cases). 
In this range, countries with the most target IPs are Brazil, Vietnam, and China, in order. The countries with dropzones in this range are European countries, including Italy, France, and the Netherlands. 
According to Table~\ref{tab:country}, a large number of dropzones exist in the US, but they also have target IPs in Brazil, Vietnam, and China, with a distance between dropzone and target in the range of 12K-14K km and 10K-12K km.

\BfPara{Country-level Analysis}
In this part, we look at the overall attack trend between dropzones and their targets on a world-scale. For each dropzone, we collect all of the target IP addresses and extract location information (\eg latitude, longitude) to display the \textit{average} position of the target area (not the exact position). The target areas are scaled according to the number of target IP addresses they contain. Fig.~\ref{fig:trend_attack} shows the results of our country-level analysis, where we limit to only plotting dropzones with more than 500 target IP addresses.
The locations of the dropzones (depicted in orange) are spread around various countries, but we highlight that there is a large concentration of target areas focused in Central Asia.

Table~\ref{tab:country} lists the top 5 countries by the number of dropzone and target IPs. 
Note that the US has a large distribution of dropzones pointing to targets in Asian countries such as Vietnam. Additionally, China and Brazil contain a large number of target IP addresses originating from European countries. To help explain these findings, we turn to the results from researchers at Kaspersky Labs. They stated that China and Vietnam were in their top-3 countries with the most-attacked IoT devices, with Brazil following closely behind~\cite{kasperskyIoT}. Imperva Incapsula (a global content delivery network and DDoS mitigation company) also confirms that Vietnam (12.8\%), Brazil (11.8\%), and China (8.8\%) were the countries with the most infected devices (from the Mirai botnet)~\cite{impervaMirai}. Moreover, these countries should intuitively contain the highest representation of vulnerable devices, such as devices with default credentials or known vulnerabilities. To validate the former, we query ``default password'' in Shodan; we found Taiwan, United States, China, Vietnam, and Thailand in the top five countries, which is partly counter-intuitive.

\begin{table}[t]
\small
\caption{Top 5 countries by the number of target and dropzone IPs. {\upshape Countries include: United States (US), Netherlands (NL), France (FR), United Kingdom (GB), Italy (IT), Vietnam (VN), Brazil (BR), China (CN), India (IN), and Pakistan (PK).}}
\label{tab:country}
\setlength{\tabcolsep}{0.35em}
\centering
\begin{tabular}{|c!{\vline width 1pt}c|c|c!{\vline width 2pt}c!{\vline width 1pt}c|c|c|}
\Xhline{2\arrayrulewidth}
\footnotesize Rank & \footnotesize Country & \scriptsize Dropzones & \footnotesize \% & \footnotesize Rank & \footnotesize Country & \scriptsize Targets & \footnotesize \% \\ \Xhline{2\arrayrulewidth}
1 & US & 1,041 & 43.25 & 1 & VN & 26,290 & 24.70 \\ \hline
2 & NL & 278 & 11.55 & 2 & BR & 20,572 & 19.33 \\ \hline
3 & FR & 188 & 7.81 & 3 & CN & 15,799 & 14.84 \\ \hline
4 & GB & 183 & 7.60 & 4 & IN & 5,598 & 5.26 \\ \hline
5 & IT & 177 & 7.35 & 5 & PK & 5,076 & 4.77 \\ 
\Xhline{2\arrayrulewidth}
\end{tabular}
\end{table}

\begin{table}[t!]
\caption{Top 5 US states by the number of target IPs and dropzone IPs. {\upshape States include: Washington (WA), New Jersey (NJ), Missouri (MO), New York (NY), Arizona (AZ), Florida (FL), New Mexico (NM), California (CA), Illinois (IL), and Michigan (MI).}}
\small
\label{tab:region}
\setlength{\tabcolsep}{0.5em}
\centering
\begin{tabular}{|c!{\vline width 1pt}c|c|c!{\vline width 2pt}c!{\vline width 1pt}c|c|c|}
\Xhline{2\arrayrulewidth}
\footnotesize Rank &\footnotesize State & \scriptsize Dropzones & \footnotesize \% & \footnotesize Rank &\footnotesize State & \scriptsize Targets & \footnotesize \%\\ 
\Xhline{2\arrayrulewidth}
1 & WA & 253 & 24.40 & 1 & FL & 506 & 30.67 \\ \hline
2 & NJ & 188 & 18.13 & 2 & NM & 356 & 21.58 \\ \hline
3 & MO & 151 & 14.56 & 3 & CA & 283 & 17.15 \\ \hline
4 & NY & 112 & 10.80 & 4 & IL & 151 & 9.15  \\ \hline
5 & AZ & 79  & 7.62  & 5 & MI & 83  & 5.03  \\
\Xhline{2\arrayrulewidth}
\end{tabular}
\end{table}

\BfPara{Region-level Analysis}
Using regional information from IPinfo~\cite{ipinfo}, we plot a heatmap representing the distribution of dropzones and targets for the entire United States. As shown in Fig.~\ref{fig:Heatmap_region_dz}, we see that Washington and New Jersey contain a high concentration of dropzones. Interestingly, Washington and New York have lots of data centers~\cite{datacenterByLocation}. We discuss this observation further in \textsection\ref{subsec:networkAnalysis}. 
Likewise, we see in Fig.~\ref{fig:Heatmap_region_target} that a high number of target IPs reside in Florida and New Mexico. Table~\ref{tab:region} lists the detailed breakdown of the top 5 dropzone and target IPs according to their US State. Overall, we had 1,037 dropzone IPs distributed over 20 US States and 1,650 target IPs spread over 22 US States.

\begin{figure}[h]
\centering
\begin{subfigure}[Distribution of dropzones by city.
\label{fig:Heatmap_city_dz}]{\includegraphics[width=0.4\textwidth]{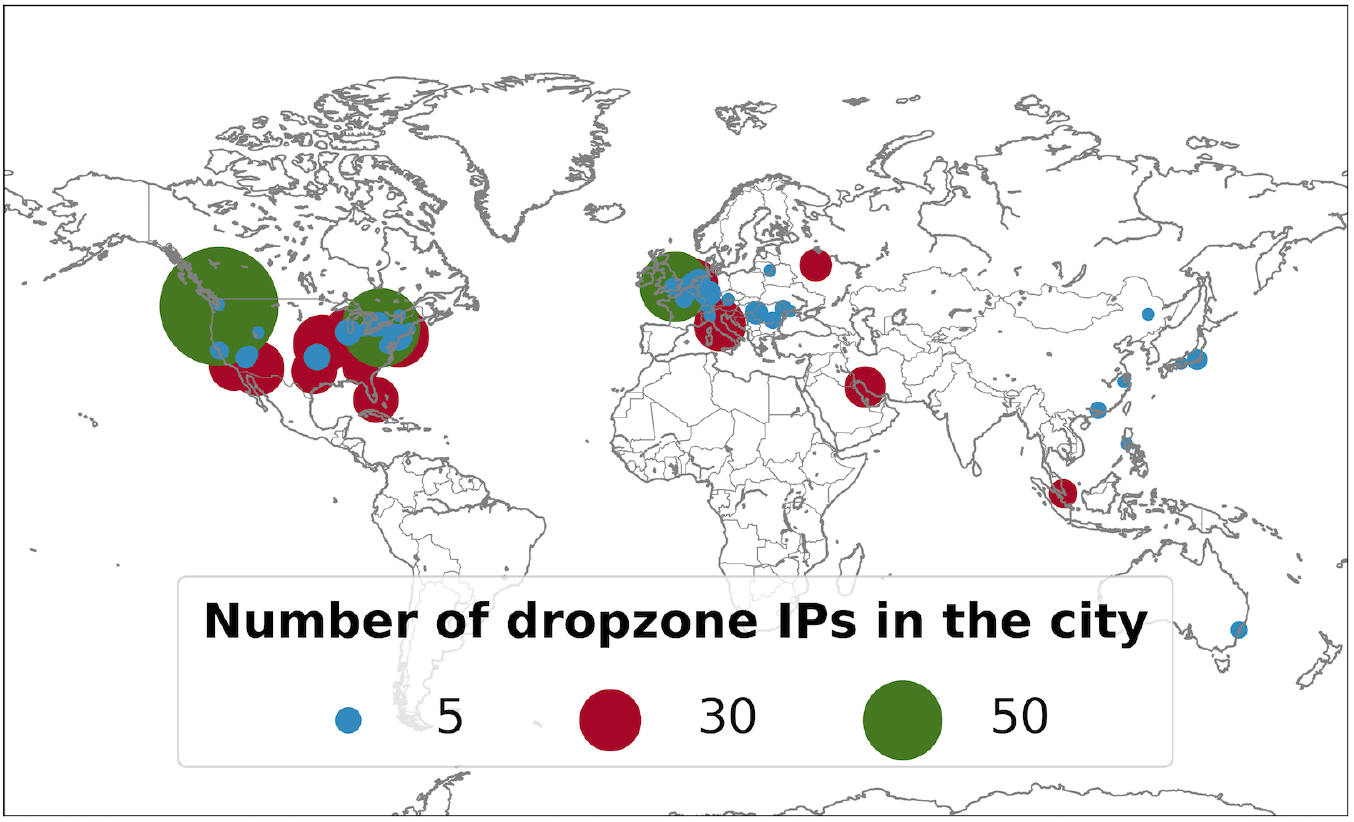}}\vspace{-2mm}
\end{subfigure}
\begin{subfigure}[Distribution of targets by city.
\label{fig:Heatmap_city_target}]{\includegraphics[width=0.4\textwidth]{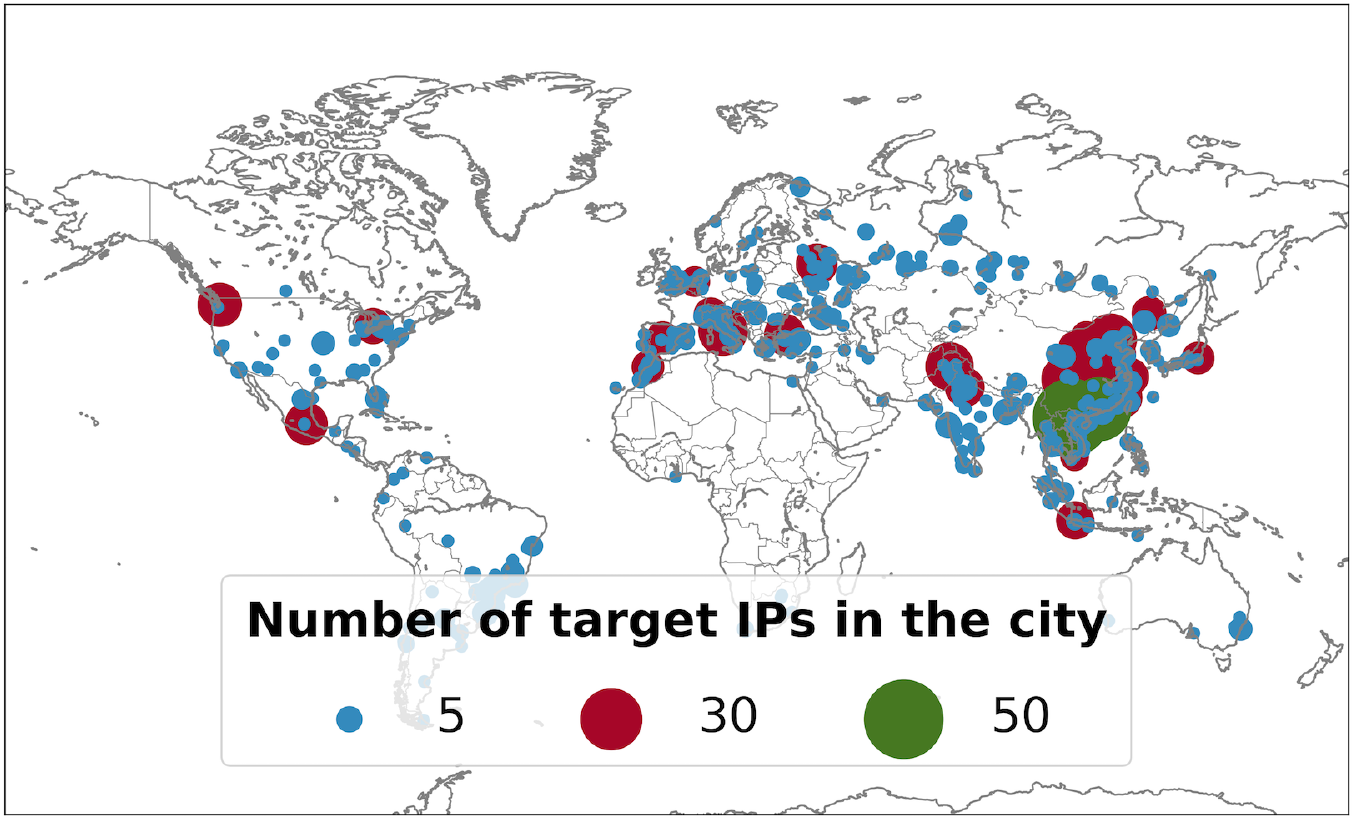}}\vspace{-2mm}
\end{subfigure}
\label{fig:Heatmap_city}
\caption{Distribution of dropzones and target IPs throughout the world. Notice that a large number of dropzones are distributed in the US and Europe and targets are mainly distributed in Asia (Vietnam, China).}
\end{figure}

\BfPara{City-level Analysis}
Using Shodan~\cite{shodan}, we look up the city in which given dropzone IPs are used, and we use IPinfo to find the city information of the target IPs  (whenever available).
We note that the city information may not exist for every IP in our data, so our region-level and city-level analyses show different distributions. Overall, we had 541 dropzone IPs distributed among 75 cities and 1,003 target IPs spread over 364 cities. In Table~\ref{tab:city}, we list the top 5 cities per the number of dropzone and target IPs. We can see the US cities top the rank for the dropzone, and China and Vietnamese cities top the ranks for the targets. 
In Fig.~\ref{fig:Heatmap_city_dz}, we note that the blue circles represent the number of dropzone IPs in the range (0, 5] with red and green circles representing dropzone IPs in ranges (5, 30] and (30, 120), respectively. Similarly, Fig.~\ref{fig:Heatmap_city_target} has the blue circle representing target IPs in the range (0, 5] with red circles as (5, 30] and green circles as (30, 50). 

\BfPara{Summary} We observe that the US has a large distribution of dropzones targeting Asian countries, e.g., Vietnam. We also see that China and Brazil are victims of sources from European countries. Kaspersky Labs'~\cite{kasperskyIoT} report and Imperva Incapsula~\cite{impervaMirai} support our findings, confirming that the Mirai botnet mostly targets Vietnam, Brazil, and China. We also observe that the most targeted cities are in the Asian countries, which are intuitive in light of the country-level results (results omitted for the lack of space). In our region and city-level analyses, we found a clear relationship between the distribution of dropzones and organizations, as shown in \textsection\ref{subsec:networkAnalysis}.

\subsection{Network Penetration Analysis}\label{subsec:networkAnalysis}
In this section, we focus on the additional attributes contained in the IP address, leveraging the information gathered from Shodan and Censys~\cite{censys}, including active ports, vulnerabilities, and organizational information.

\begin{figure}[t!]
\centering
\begin{subfigure}[Distribution of dropzones by US State.
\label{fig:Heatmap_region_dz}]{\includegraphics[width=0.7\linewidth]{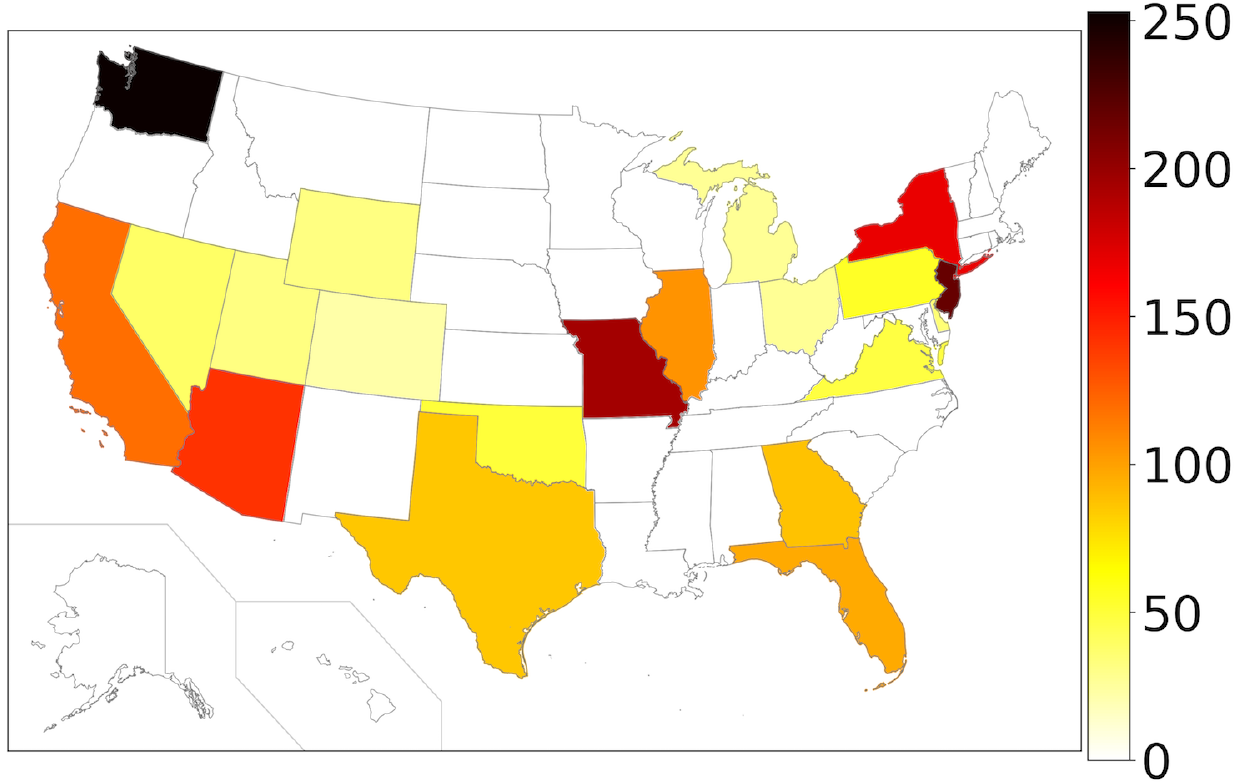}}
\end{subfigure}

\begin{subfigure}[Distribution of targets by US State.
\label{fig:Heatmap_region_target}]{\includegraphics[width=0.7\linewidth]{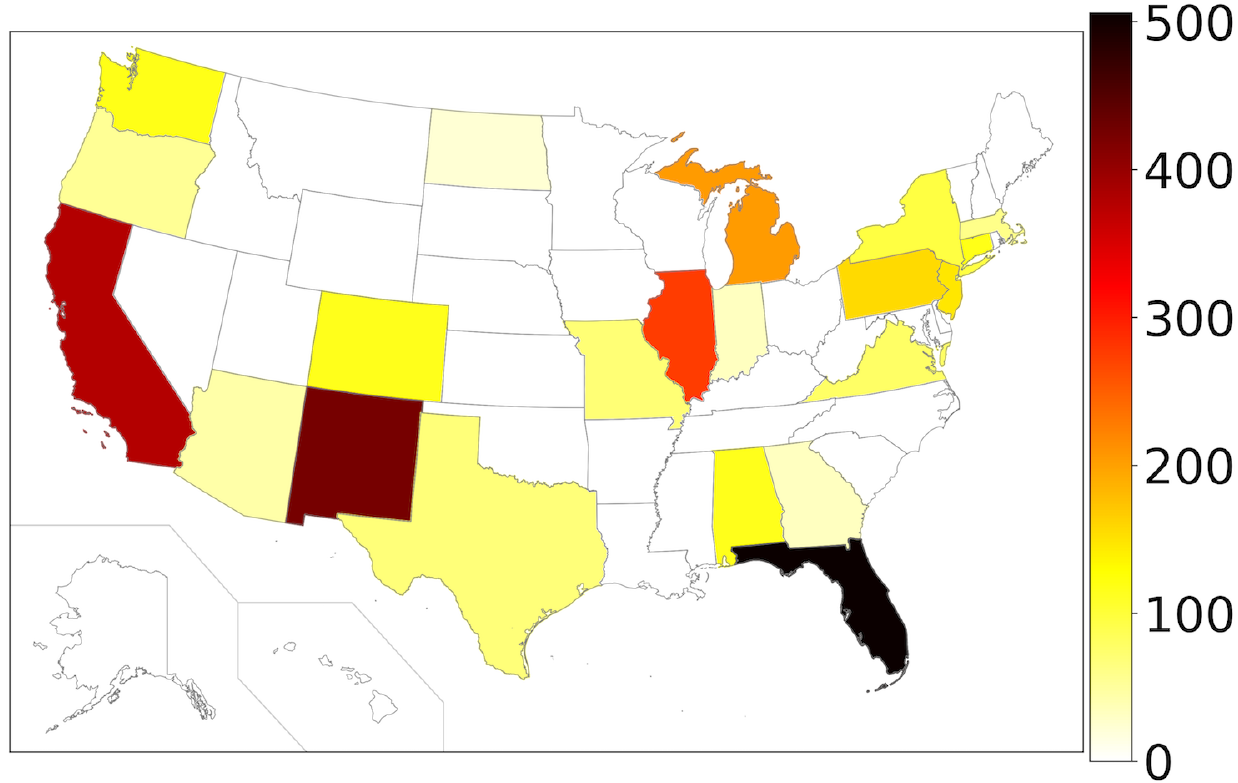}}
\end{subfigure}

\label{fig:Heatmap_region}
\caption{Distribution of dropzones and target IPs in the United States. This figure shows dropzone and target mainly exist in which state in the US.}
\end{figure}

\begin{table}[t]
\caption{Top 5 Cities per dropzone and target IPs. Cities in the US occupy top ranks with respect to dropzone IPs and cities in Vietnam and China have a lot of target IPs, which demonstrate a case similar to the country-level analysis.}
\label{tab:city}
\centering
\setlength{\tabcolsep}{0.5em}
\scriptsize
\begin{tabular}{|c!{\vline width 1pt}l|c|c!{\vline width 2pt}c!{\vline width 1pt}l|c|c|}
\Xhline{2\arrayrulewidth}
Rank & City &  Dropzones & \% & Rank & City &  Targets & \% \\ 
\Xhline{2\arrayrulewidth}
1 & Seattle     & 113 & 20.89 & 1 & Hanoi           & 48 & 4.79 \\ \hline
2 & Buffalo     & 49  & 9.06  & 2 & Guangzhou       & 32  & 3.19  \\ \hline
3 & London      & 39  & 7.21  & 3 & Beijing         & 21  & 2.09  \\ \hline
4 & Clifton     & 29  & 5.36  & 4 & Rome            & 19  & 1.89  \\ \hline
5 & Kansas & 27  & 4.99  & 5 & Islamabad       & 18  & 1.79  \\
\Xhline{2\arrayrulewidth}
\end{tabular}
\end{table}

\BfPara{Active Ports}
For each dropzone and target IP address, we obtain a list of active ports from Shodan and Censys. In total, we extracted 5,745 active ports from 716 of the 877 dropzone IPs and 1,114 active ports from 129 of 189 of the non-masked target IPs. 
We note that malware authors mask the octets of the target IP addresses, which they determine dynamically at runtime. In this analysis, we only use the IP addresses that are not masked. A summary of our results is shown in Table~\ref{tab:ports_dz} and Table~\ref{tab:ports_target}. We notice that the largest portion of active ports is common services such as SSH (port 22), HTTP (port 80), and HTTPS (port 443). However, other active ports associated with target IPs are highlighted in Table~\ref{tab:ports_target}, including the SUN Remote Procedure Call (RPC) on port 111 and the Network Time Protocol (NTP) on port 123.

Port 111 is used by the \textit{Port Mapper} service over the TCP and UDP protocols~\cite{rfc1833}, which essentially is a port lookup service for the Open Network Computing Remote Procedure Call (ONC RPC) system designed by Sun Microsystems in the 1980s for their Network File System~\cite{speedguide111}. As described in RFC 1833~\cite{rfc1833}, the port numbers for RPC programs and services are determined dynamically on startup, so if a client wishes to make ONC RPC calls, they will query the Port Mapper on port 111 to obtain the appropriate RPC service port. As reported by L3 Communications in August  2015~\cite{l3comm}, the Port Mapper service became a new attack vector for adversaries seeking to amplify their DDoS attacks. This is due to the fact that when Port Mapper is queried, the response size varies significantly depending on which RPC services are available on the host. In their examples, L3 Communications show that a 68-byte query results in a 486-byte response for an amplification factor of 7.1x with responses as large as 1,930 bytes for amplification of 28.4x. If adversaries spoof the victim's  IP for UDP packets directed towards vulnerable devices with port 111 open, they will ultimately be redirected back \textit{en masse} towards the victim (\ie a UDP flood attack).

Port 123 is reserved for the Network Time Protocol (NTP)~\cite{rfc958}, which is used to synchronize network clocks using a set of distributed clients and servers. Per various reports, NTP could be abused in DDoS amplification attacks~\cite{graham-cumming2013}. Much like the exploit for the Port Mapper service described above, NTP is also UDP-based and can be prone to ``IP spoofing'' for DDoS attacks~\cite{graham-cumming2014}. As emphasized in~\cite{verisignUdpFlood}, exploiting NTP has a great potential for amplification attacks due to the ``monlist'' command that a typical attacker sends to NTP servers, which returns the last 600 IP addresses previously synchronized with the NTP server using 30 separate UDP packets, each of which is 448 bytes. The overall size varies depending on the server, but the data volume is almost 1,000x larger than the packet originally sent by the attacker.

\begin{table}[t]
\caption{Top 10 active ports in dropzone IPs. Most shown ports are used for well-known network services.}
\small
\label{tab:ports_dz}
\centering
\setlength{\tabcolsep}{0.25em}
\begin{tabular}{|c!{\vline width 1pt}c|c|c|l|l|}
\Xhline{2\arrayrulewidth}
Rank & Port & Count & \% & Service &  Description                 \\ \Xhline{2\arrayrulewidth}
1 & 22 & 641 & 32.57 & {\scriptsize SSH} & {\scriptsize The Secure Shell (SSH) Protocol} \\ \hline
2 & 80 & 600 & 30.49 & {\scriptsize HTTP} & {\scriptsize World Wide Web HTTP} \\ \hline
3 & 443 & 350 & 17.78 & {\scriptsize HTTPS} & {\scriptsize HTTP protocol over TLS/SSL} \\ \hline
4 & 25 & 276 & 14.02 & {\scriptsize SMTP} & {\scriptsize Simple Mail Transfer} \\ \hline
5 & 21 & 275 & 13.97 & {\scriptsize FTP} & {\scriptsize File Transfer Protocol {[}Control{]}} \\ \hline
6 & 3306 & 224 & 11.38 & {\scriptsize MySQL} & {\scriptsize MySQL database system} \\ \hline
7 & 53 & 187 & 9.50 & {\scriptsize DNS} & {\scriptsize Domain Name Server} \\ \hline
8 & 110 & 175 & 8.89 & {\scriptsize POP3} & {\scriptsize Post Office Protocol - Version 3} \\ \hline
9 & 143 & 171 & 8.69 & {\scriptsize IMAP} & {\scriptsize Internet Message Access Protocol} \\ \hline
10 & 993 & 165 & 8.38 & {\scriptsize IMAPS} & {\scriptsize IMAP over TLS protocol} \\ 
\Xhline{2\arrayrulewidth}
\end{tabular}
\end{table}

\BfPara{Vulnerabilities} We then explore the susceptibility of IP addresses by examining the vulnerabilities associated with services running on them. 
For the vulnerable endpoints, we gather the Common Vulnerabilities and Exposures identifier (CVE-ID), a unique identifier assigned by MITRE~\cite{mitreCve} to standardize security vulnerabilities.

We analyze the vulnerabilities in the dropzones to understand their dynamics. Table~\ref{tab:vul_dz} shows the top six vulnerabilities by the number of dropzone IPs, and we further analyze them to assess their root causes. We found that CVE-2017-15906 is the most frequent, which is found in 203 dropzone IPs, affecting 448 instances of malware. According to the National Vulnerability Database (NVD)~\cite{nvd}, CVE-2017-15906 is a ``medium'' severity vulnerability where versions of OpenSSH before 7.6 do not properly prevent write operations in \textit{readonly} mode, allowing attackers to create several zero-length files that could exhaust disk space. The second was CVE-2014-1692, labeled by NVD as ``high'' severity vulnerability, where it allows remote attackers to launch a DoS attack through memory corruption due to uninitialized data structures from the \texttt{hash\_buffer} function in OpenSSH. We note that these vulnerabilities are prevalent to the target devices and do not tell much about the impacted dropzones. We observe that 98.61\% of dropzone IPs host a service that has the vulnerability CVE-2014-1692. Additionally, one or more of the vulnerabilities facilitate unauthorized authentication, rendering device-level access. These observations reveal that a remote unauthorized authentication vulnerability can be an indicator of potential dropzone.

\begin{table}[t]
\caption{Top 10 active ports in target IPs. With the exception of a few, most shown ports are used for common services.}
\small
\label{tab:ports_target}
\centering
\setlength{\tabcolsep}{0.25em}
\begin{tabular}{|c!{\vline width 1pt}c|c|c|l|l|}
\Xhline{2\arrayrulewidth}
Rank & Port & Count & \% & Service & Description                 \\ 
\Xhline{2\arrayrulewidth}
1 & 80 & 111 & 17.85 & {\scriptsize HTTP} & {\scriptsize World Wide Web HTTP} \\ \hline
2 & 22 & 106 & 17.04 & {\scriptsize SSH} & {\scriptsize The Secure Shell (SSH) Protocol} \\ \hline
3 & 443 & 67 & 10.77 & {\scriptsize HTTPS} & {\scriptsize HTTP protocol over TLS/SSL} \\ \hline
4 & 21 & 51 & 8.20 & {\scriptsize FTP} & {\scriptsize File Transfer Protocol {[}Control{]}} \\ \hline
5 & 25 & 49 & 7.88 & {\scriptsize SMTP} & {\scriptsize Simple Mail Transfer} \\ \hline
6 & 3306 & 40 & 6.43 & {\scriptsize MySQL} & {\scriptsize MySQL database system} \\ \hline
7 & 53 & 29 & 4.66 & {\scriptsize DNS} & {\scriptsize Domain Name Server} \\ \hline
8 & 8080 & 29 & 4.66 & {\scriptsize HTTP-alt} & {\scriptsize HTTP Alternate (see port 80)} \\ \hline
9 & 111 & 28 & 4.50 & {\scriptsize SunRPC} & {\scriptsize SUN Remote Procedure Call} \\ \hline
10 & 123 & 26 & 4.18 & {\scriptsize NTP} & {\scriptsize Network Time Protocol} \\
\Xhline{2\arrayrulewidth}
\end{tabular}
\end{table}

We also observe vulnerabilities that result in bypassing authentication and crossing privilege boundaries, including CVE-2016-0777, CVE-2012-0814, and CVE-2010-4478, which are vulnerabilities that allow an attacker to obtain access permission on target devices by stealing sensitive information; (\eg private key or authorized key). For example, we observed that CVE-2011-4237 enables remote attackers to make an unauthorized modification to the list of authenticated keys by injecting an arbitrary HTTP header. More specifically, we observe that 17.04\% of dropzone IP addresses (\ie 144) have at least one of these four vulnerabilities. 
These vulnerabilities provide a broad range of attack capabilities that can be used by attackers to compromise the devices, then act as dropzones. Moreover, where some of the dropzones use the default credentials make the devices an attractive target. 

\begin{table}[t]
\caption{Top 6 Vulnerabilities by the number of dropzone IPs. Note that the dropzones use vulnerable versions of OpenSSH.}
\label{tab:vul_dz}
\centering
\setlength{\tabcolsep}{0.25em}
\small
\begin{tabular}{|l!{\vline width 1pt}c|c|l|}
\Xhline{2\arrayrulewidth}
Vulnerability & IP & Malware & Description\\ \Xhline{2\arrayrulewidth}
CVE-2017-15906 & 203 & 448 & {\scriptsize OpenSSH/DDoS} \\ \hline
CVE-2014-1692 & 142 & 320 & {\scriptsize OpenSSH/DDoS} \\ \hline
CVE-2016-0777 & 142 & 325 & {\scriptsize OpenSSH/Private Key leakage} \\ \hline
CVE-2012-0814 & 140 & 307 & {\scriptsize Cross-privilege boundaries/OpenSSH} \\ \hline
CVE-2011-4327 & 140 & 307 & {\scriptsize OpenSSH/Authentication leakage} \\ \hline
CVE-2010-4478 & 140 & 307 & {\scriptsize OpenSSH/Authentication override} \\ \Xhline{2\arrayrulewidth}
\end{tabular}
\end{table}

\begin{figure*}[t]
\centering
\begin{subfigure}[Organizations v. the number of dropzone IP.
\label{fig:CDF_org_DZ_IP}]{\includegraphics[width=0.3\linewidth]{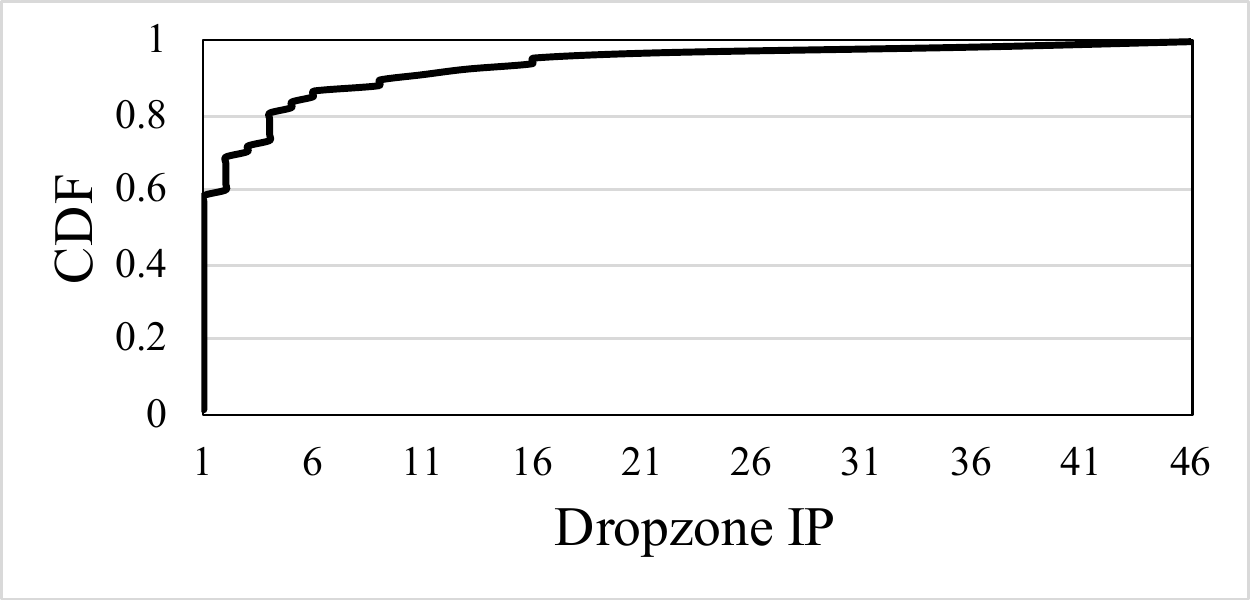}}
\end{subfigure}
\begin{subfigure}[Organization v. the number of malware.
\label{fig:CDF_org_Malware}]{\includegraphics[width=0.3\linewidth]{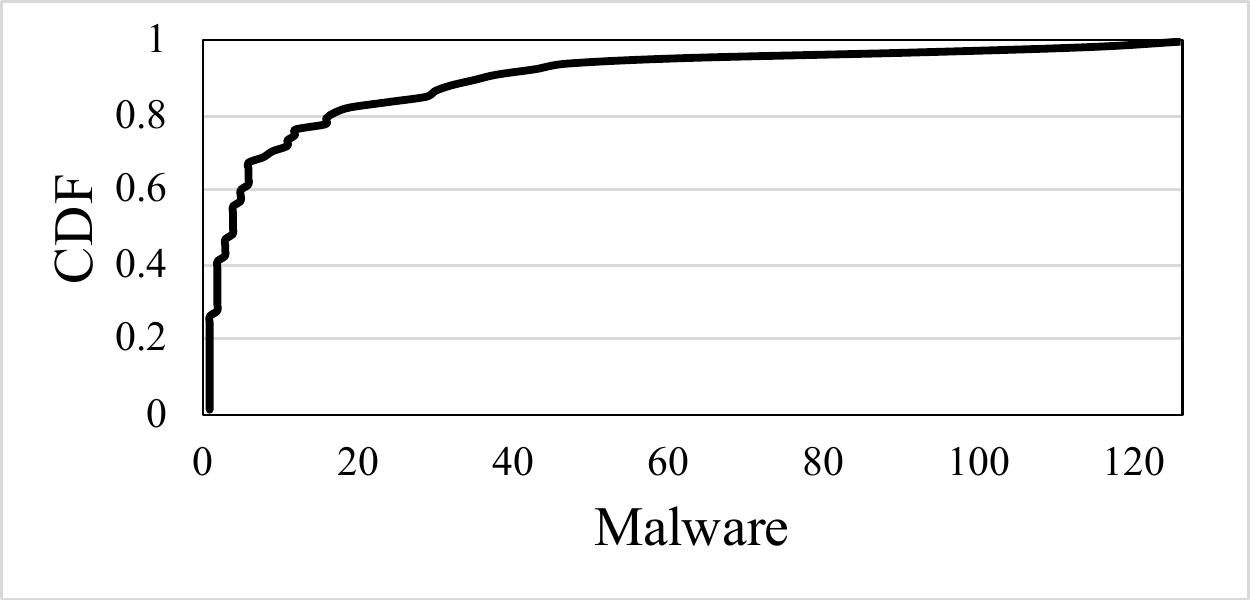}}
\end{subfigure}
\begin{subfigure}[Organization v. the number of target IP.
\label{fig:CDF_org_Target_IP}]{\includegraphics[width=0.3\linewidth]{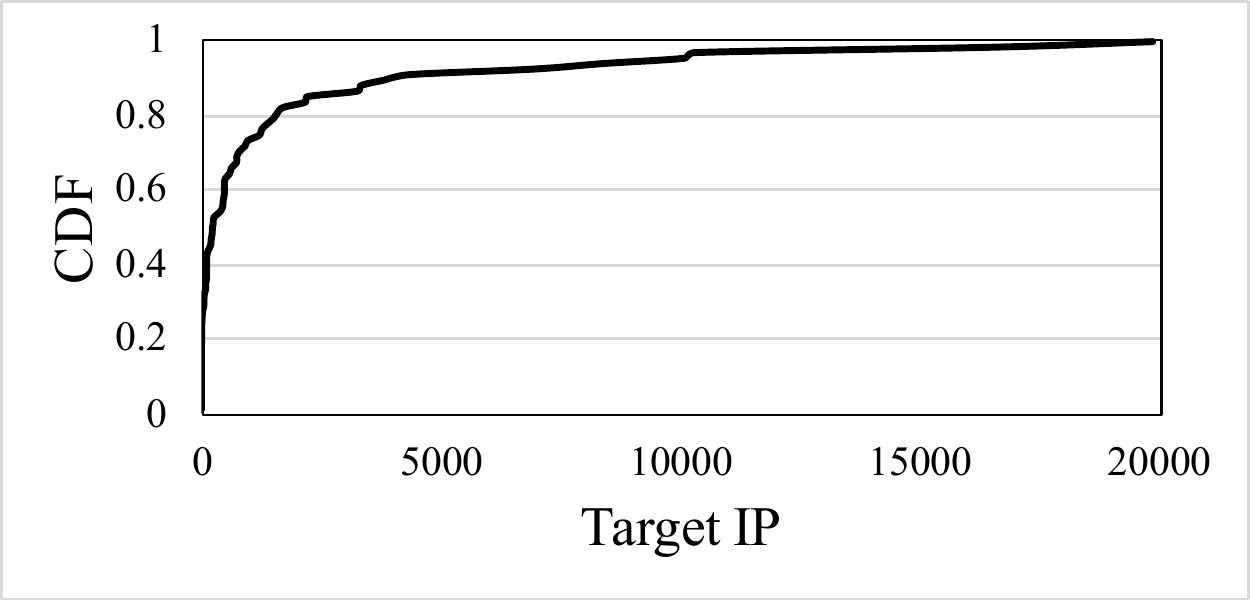}}
\end{subfigure}
\caption{Organization v. the number of dropzone IP, malware and target IP. The similarity in the CDFs is because there is a high probability that the organization that includes more malware will contain more dropzone and target, except in a few cases (\ie where one dropzone refers to thousands of targets, or one dropzone refers to only one target).}
\label{fig:CDF_org} 
\end{figure*}

\BfPara{IP-Owning Organizations} We now examine the organizations that own the given IP spaces identified in our analysis. In Fig.~\ref{fig:CDF_org} we provide plots for the CDF contrasting the IP-owning organizations and the number of dropzone IPs, number of malware instances, and number of target IPs they point to. For statistical instances of the head of the distribution, a breakdown of the top 10 IP-owning organizations is presented in Table~\ref{tab:org_dz}. 
We notice that for each organization, there is a clear relationship between the number of dropzone IPs and the number of malware instances they are associated with. However, there are abnormal cases in our dataset, such as Cogeco Peer one (Canada) and MAROSNET  Telecommunication Company LLC (Russia), which have only one dropzone IP but point to 2,214 and 2,178 target IPs, respectively. In contrast, HOSTKEY (that operates in the Netherlands and Russia~\cite{hostkey}) has only two dropzone IPs that are associated with two target IPs and two instances of malware. 

Interestingly enough, the \textit{locations of these organizations coincide with the heatmap of US States} presented in Fig.~\ref{fig:Heatmap_region_dz}, which illustrates the highest distribution of dropzones. For example, the organization with the greatest number of dropzone IPs according to our data is Wowrack.com, which is a cloud service provider with headquarter offices in Seattle, Washington~\cite{wowrackAbout}. Besides, Wowrack operates eight other data centers in multiple cities across the United States and Southeast Asia. 
As reported by AbuseIPDB~\cite{abuseipdb}, Wowrack.com IPs have received several complaints of abusive activity from multiple sources. The Canadian Internet Registration Authority (CIRA) has also urged administrators to block domains originating from Wowrack.com (\eg \texttt{ns6.wowrack.com}) because they are associated with the Mirai IoT botnet~\cite{Williamson2018}.

By referring to Fig.~\ref{fig:Heatmap_region_dz}, we also notice that New York State contains a high number of dropzones-{}-which is most likely caused by two organizations from Table~\ref{tab:org_dz} that have data centers in the city of Buffalo, NY (Wowrack.com and ColoCrossing~\cite{colocrossing}). Also highlighted in red is the state of Arizona, which is home of Input Output Flood LLC~\cite{ioflood} (ranked 10\textsuperscript{th} for organizations with several dropzones).

We cannot say for sure why these organizations contain the most dropzone IPs, although one can speculate that they are more tolerant to harboring customers who engage in malicious activities, or are subject to compromise. For example, the organization with the 3\textsuperscript{rd}-highest number of dropzones in our dataset is Choopa LLC, with a primary Point of Presence (POP) in the State of New Jersey~\cite{choopa}, which one can clearly see on the heatmap (colored dark-red) shown in Fig.~\ref{fig:Heatmap_region_dz}. While online public reviews are not an authoritative source for quality and security~\cite{Tuttle2012}, they are useful in shedding light on this organization: the low ratings from the Google reviews of Choopa LLC put them in a negative light. 

\begin{table}[t]
\caption{Top 10 organizations by the number of their dropzone IPs. In this table, most organizations increase linearly in the number of dropzone, malware, and target.} 
\small
\label{tab:org_dz}
\centering
\setlength{\tabcolsep}{0.5em}
\begin{tabular}{|c!{\vline width 1pt}c|c|c|l|}
\Xhline{2\arrayrulewidth}
\scriptsize Rank & \scriptsize \# Dropzone & \scriptsize \# Target & \scriptsize \# Malware & {\footnotesize Organization}       \\ \Xhline{2\arrayrulewidth}
1 & 46 & 16,586 & 126 & {\footnotesize Wowrack.com} \\ \hline
2 & 36 & 19,878 & 114 & {\footnotesize Aruba S.p.A} \\ \hline
3 & 22 & 10,282 & 62 & {\footnotesize Choopa LLC} \\ \hline
4 & 16 & 3,816 & 47 & {\footnotesize DigitalOcean LLC} \\ \hline
5 & 16 & 3,330 & 29 & {\footnotesize ColoCrossing } \\ \hline
6 & 13 & 8,373 & 38 & {\footnotesize NForce Entertainment B.V.} \\ \hline
7 & 11 & 1,701 & 24 & {\footnotesize Hydra Communications Ltd} \\ \hline
8 & 9 & 4,354 & 35 & {\footnotesize Ad Net Market Media Srl} \\ \hline
9 & 9 & 388 & 17 & {\footnotesize Wholesale Data Center LLC} \\ \hline
10 & 6 & 1,220 & 8 & {\footnotesize Input Output Flood LLC} \\ 
\Xhline{2\arrayrulewidth}
\end{tabular} 
\end{table}

\BfPara{Summary} We observe that while ports 80 and 22 are the most widely used across endpoints, the usage of port 111 and port 123 by the target IP addresses is predominant. Being two possible entry points for attacker, the usage should be curtailed. We also observe the low presence of vulnerable services by dropzone IP addresses, where vulnerabilities in the dropzones indicate that the attackers override their authentication status and then utilize the OpenSSH vulnerabilities to gain access to the device. 
Our analysis also indicates a tolerance of organizations towards the endpoints used in malware; the divergence between monetary profit and trust loss among users (prospective domain buyers) deserves further investigation.

\section{Network centric analysis}\label{sec:networkCentric}

Malware specifically aimed at IoT devices tend to recruit a large number of intermediary targets to launch attacks on high-profile targets ultimately. The malware typically identifies the intermediary targets using their IP addresses which are either mentioned in their source code or downloaded via dropzone. For example, Fig.~\ref{fig:target_list} demonstrates how malware attempts to download a file named {\em DNS.txt} which possibly contains the list of target IP addresses. As for the endpoints in the code base, these could include IP addresses and masked IP addresses, only a prefix (\eg \texttt{123.17.\%d.\%d}).

In the previous sections, we analyzed the IP addresses explicitly in the malware code base. For masked IP addresses, malware typically uses functions to hide the targets from the malware analysts and determine the targets dynamically. These functions are invoked during run time to determine the remaining of the masked octets. Malware authors seldom obfuscate these functions -- we, therefore, in this section, examine the entire /16, /24, or /8 network to probe their susceptibility.

\begin{figure}[t!]
	\centering
	\includegraphics[width=.99\linewidth]{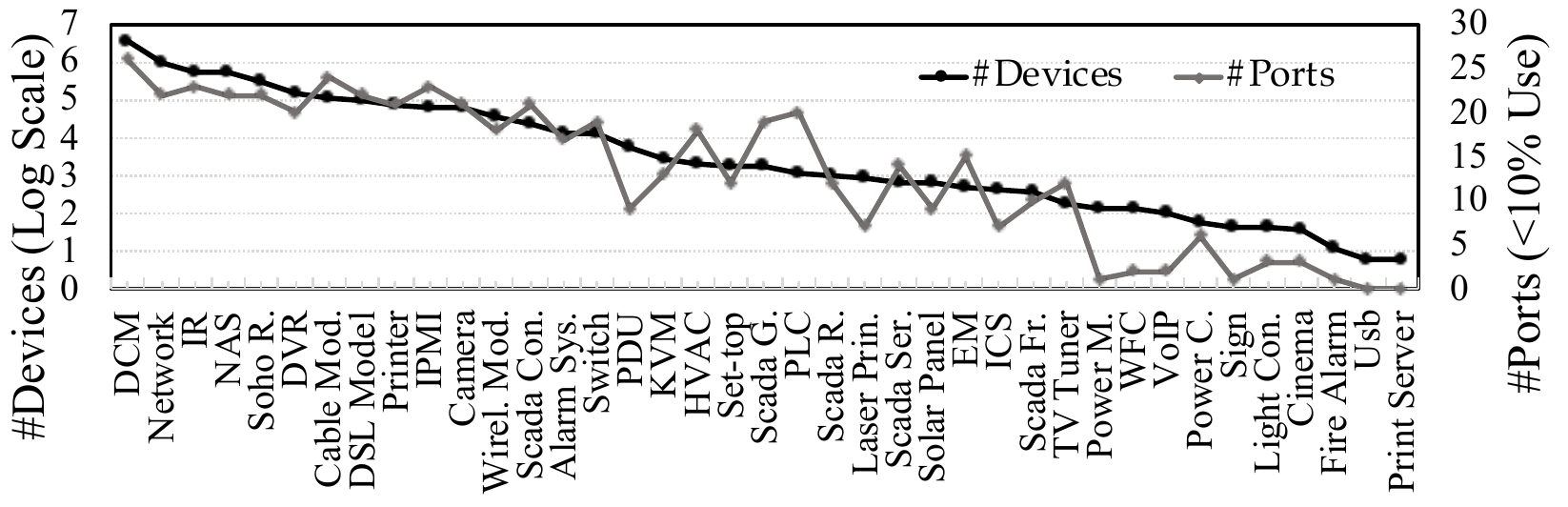} 
	\caption{Total number of devices and ports that used less than 10\% of devices. The left Y-axis is the number of ports, the right Y-axis is the total number of devices, and the X-axis is the device types. Device types include: DSL/cable Modem (DCM), Infrastructure Router (IR), Network Attached Storage (NAS), Digital Video Recorder (DVR), Intelligent Platform Management Interface, (IPMI) Power Distribution Unit (PDU), Kernel-based Virtual Machine (KVM), Heating, ventilation, and Air Conditioning (HVAC), Programmable Logic Controller (PLC), Environment Monitor (EM), Industrial Control System (ICS), and Water Flow Controller (WFC).}
	\label{fig:twin_port_total_devices} 
\end{figure}

Using CIDR notation, Table~\ref{tab:tg_composition} shows that 98.92\% of the target endpoints are masked, mapping to 126 unique /8 networks and 1,869 unique /16 networks and 27 unique /24 networks. Removing the /16 networks covered in /8 and /24 networks, we have 125 /8 networks and 435 /16 networks. We evaluated and analyzed these networks to investigate their exposure to risk. In particular, we examined the devices on these networks and looked at the services being used.
These 560 networks are then searched on Censys~\cite{DurumericAMB15} which maps to 100,793,403 active IP addresses, which also allows us to analyze their active ports. As different devices use different services to operate, we clustered the IP addresses by their device types and studied which ports were being used. Considering that open ports lead to increased security risks, we look for ports that are necessary for a device to operate without any misfire. Taking a conservative approach, we suggest that if a port is being used by less than 10\% of devices in a given device type, it should be closed to reduce its exposure to risk. We observe that except for {\em VoIP phone} (over 77\% of them used five ports), more than 75\% of the devices among all the other device types have only two or fewer ports being used. Fig.~\ref{fig:twin_port_total_devices} shows the number of devices within a device type in log scale and the number of ports being used by less than 10\% of the devices. In this figure, the two graphs show a similar pattern. We speculate this result is due to more attacks taking place on the popular devices (\eg target devices of the Mirai consist of security cameras, DVRs, and consumer routers \cite{AntonakakisABB17}). 
Additionally, the susceptibility of high-wattage IoT devices, such as heating, ventilation, and air conditioning (HVAC), power distribution units (PDU), \etc, can be abused by the attackers to launch large-scale coordinated attacks, \eg power grids, as has been demonstrated by Soltan \etal~\cite{SoltanMP18}.

\BfPara{Summary} The division of the endpoints by devices and then determining their exposure to the attackers represent the chances of an endpoint being compromised. Based on our analysis, we suggest the users close the ports that aren't necessary for the uninterrupted execution of their devices. These endpoints need to be further examined in-depth to understand the pattern that could predict an endpoint's chances of being compromised. The suggestions could be finally narrowed, with specific device centered recommendations, and by probing them individually by performing an offensive penetration testing. However, in this work, we understand the data-centric landscape and put forward the suggestions with a conservative approach and without carrying out any offensive analysis undermining ethics.

\begin{table}[t!]
\caption{Composition of Target IPs for masked and not-masked networks. {\upshape ``In Total" means the total number of target IPs, ``In Unique" means the composition of non-duplicated target IPs.}}
\small
\label{tab:tg_composition}
\centering
\begin{tabular}{|c!{\vline width 1pt}c|c|c|c|}
\Xhline{2\arrayrulewidth}
Address & Total & \% & Unique & \% \\ \Xhline{2\arrayrulewidth}
/24 & 137 & 0.13& 27& 1.22\\ \hline
/16 & 104,369 & 98.07 & 1,869 & 84.53 \\ \hline
/8 & 776 & 0.73 & 126 & 5.70\\ \hline
\small Not-masked & 1,146 & 1.08 & 189 & 8.55\\ \hline
Total& 106,428 & 100.00 & 2,211 & 100.00 \\
\Xhline{2\arrayrulewidth}
\end{tabular} 
\end{table}

\section{Related Work}\label{sec:related}
\begin{table}[h]
\centering
\caption{Comparison with Related work. We compare the objective of the works, the analysis method used, static (S) or dynamic (D), and the dataset size.}
\label{tab:related}
\scalebox{0.9}{
\begin{tabular}{|l|l|l|l|}
\Xhline{3\arrayrulewidth}
Study                 & Objective & Method & Dataset        \\ \hline
Antonakakis~\etal~\cite{AntonakakisABB17}           & Analysis         & D    & Honeypot, Logs \\\hline
Cozzi~\etal~\cite{CozziGFB2018}                 & Analysis         & S/D  & 10,548         \\\hline
Pa~\etal~\cite{Pa2016_IoTPOT}                & Analysis         & D    & Honeypot       \\\hline
Vervier~\etal~\cite{VervierS18}               & Analysis         & D    & Honeypot       \\\hline
Anwar~\etal~\cite{AnwarAPW20}                 & Detection       & S      & 2,899          \\\hline
Bendiab~\etal~\cite{BendiabSAK20}          & Detection         & D    & 1,000          \\
\Xhline{3\arrayrulewidth}
\end{tabular}}
\end{table}

A summary of the related work is in~\autoref{tab:related}. Recent studies related to IoT malware have primarily focused on classifying IoT Malware.  IoTPOT, proposed by Pa~\etal~\cite{Pa2016_IoTPOT}, was one of the first honeypots specifically for IoT threats. Antonakakis~\etal~\cite{AntonakakisABB17} analyzed the Mirai botnet to understand their execution. Cozzi~\etal~\cite{CozziGFB2018} investigated the malware samples to understand their capabilities. Anwar~\etal~\cite{AnwarAPW20} statically analyzed the IoT malware binaries to exhibit their features and strategies. Multiple studies have proposed machine and deep learning-based detection systems using statically generated features~\cite{alasmary2018graph,alasmary2019analyzing,AnwarAPW20,SuVPS18,BendiabSAK20}. Sivanathan~\etal~\cite{SivanathanSGRWV17} further characterized the network traffic attributes. Rafique and Caballero~\cite{RafiqueC13} used the network signatures from executing malware binaries to cluster them into families. Additionally, West and Mohaisen~\cite{WestM14} used 28,000 expert-labeled endpoints extracted from $\approx$100K malware binaries for binary threat classification. Deep learning algorithms have been leveraged to identify malicious endpoints~\cite{OuellettePL13}.

Lei~\etal~\cite{LeiQWL19} proposed a graph-based event-aware malware detection tool for smart IoT devices through the event groups to exploit their Android application's behavioral patterns. Vervier and Yun~\cite{YunV19} propose high-level security and privacy labels for IoT devices towards consumer awareness. Alrawi~\etal~\cite{AlrawiLAM19} evaluated the IoT devices deployed in the home environment. Further, Perdisci~\etal~\cite{PerdisciPAA20} proposed IoTFinder to identify IoT devices by analyzing the passive DNS traffic.

Studies have used internet-scanning services (\eg Shodan~\cite{shodan}) augmenting it with information gathered from the known vulnerability databases (\eg NVD~\cite{nvd}) to measure the potential risks of Internet-connected devices. For example, Genge and Enachescu\cite{GengeE16} proposed ShoVAT (Shodan-based Vulnerability Assessment Tool). They collected IoT device information such as open ports, when they were scanned, banner data, and their operating system through the Shodan API. Formby \etal~\cite{FormbySLR16} outlined security challenges in the existing Industrial Control Systems (ICS) and addressed them by leveraging fingerprinting methods. Feng \etal~\cite{FengLWS18} proposed rule-based discovery and annotation of IoT devices.

To the best of our knowledge, there is no recent work that analyzes the relationships between the endpoints of IoT malware dropzones and their target devices. With that said, the closest study to our work is by Vervier~\etal~\cite{VervierS18} who deployed a honeypot to understand the threat landscape and the operational pattern of the IoT malware. 
Additionally, Holz~\etal~\cite{HolzEF09} presented one of the first empirical studies of malware and dropzones conducted over a seven-month in late 2008, focusing on keystroke-collecting malware (``keyloggers'') with dynamic analysis using \textit{CWSandbox}. Since keyloggers typically contact dropzones upon execution, the authors successfully obtain the locations of several dropzones mapped to different countries, shown to be Russia and the US, among others. Although limited, prior works have looked into investigating Linux malware, and the malware endpoints have not received the required attention. Cozzi~\etal~\cite{CozziVDSBB20} analyze a dataset of IoT malware and note that the obfuscation is rare in IoT malware, thereby making static analysis worthy. 


\section{Discussion}\label{sec:discussion}
Our study focuses on the IP addresses present in the code base of the IoT malware. Overall, we analyze the dropzone and the target IP addresses, covering $\approx$78\% of the public IPv4 space. In this section, we discuss the implications of our study. 

\BfPara{Possibility of re-attack and need for region-specific defense} Our analysis shows re-usability of target IP addresses by different dropzone IP addresses, which may be due to the use of endpoints search through IoT search engines like Shodan or Censys. This exhibits chances of race among the different adversaries to get hold of the susceptible devices (\eg, use of default credentials). 

We observe that the source of the attack is concentrated in the United States and Europe, while the targeted endpoints are concentrated towards South Asia. This shows a varying security posture of the different regions. For example, despite being a small country in South Asia, Vietnam accounts for $\approx$25\% of the targeted endpoints, while Brazil is a distant second with $\approx$19\% targeted endpoints. This warrants a region-specific approach towards defense. 

The study of open ports per device type shows the high presence of open ports. We suggest that the ports that are not being used by 90\% of the devices be closed are unnecessary, and focusing on manufacturers and their service requirement would help better understand the targeted devices.

\BfPara{Patch prioritization} Vulnerability analysis points at the existing endpoints with vulnerable services running on them. However, it is common knowledge that organizations and device manufacturers prioritize the patching of vulnerabilities, considering the open-source and wide reporting of vulnerabilities. When prioritizing, they make use of vulnerability scoring systems, such as the Common Vulnerability Scoring System (CVSS). CVSS version 3 assigns the vulnerabilities a severity level of low, medium, high, or critical depending on the vulnerability characteristics, such as impact. 

A patch prioritization employment in an organization would push the vulnerabilities with critical severity at the top of the priority and those with low severity to the bottom. However, non of the vulnerabilities in ~\autoref{tab:vul_dz} have a critical severity. We, therefore, suggest making use of vulnerability exploited by malware as a modality for patch prioritization.

\BfPara{Validity} Given the bounded size of our dataset, one may argue that our dataset and associated analyses do not necessarily give a complete view of the dynamics of the IoT malware endpoints. However, we argue that this view is representative, and support this argument with an updated experiment. 

To examine the validity of the dataset used and the IP addresses analyzed for this study, we contrast the persistence of the dropzone and target IP addresses outside the scope of the dataset used in this study. We observe a continued use of the dropzone addresses analyzed in this study, reflecting the persistence of the identified attack sources. Recall that we identify 877 unique dropzone IP addresses, 436 (49.7\%) of which are observed in the dataset used for comparison. 

Overall, the IP addresses observed in the study are present in 35.7\% (2,251 out of 6,303) of the malware in the dataset used for comparison. Additionally, we investigate the usage of the vulnerable services leading to the most frequent vulnerabilities by the devices on the public Internet. We observe that the vulnerable services are still in use, reflecting the possible use of these devices in future malware campaigns. Moreover, we cover 78.2\% of the public IPv4 space, exhibiting the nature of the reachable devices.

\section{Concluding Remarks}\label{sec:conclusion}

In this paper, we analyze the $\approx$78.2\% of total responsive public IPv4 endpoints among dropzones and their targets as extracted from IoT software and spread across the globe from diverse perspectives.
First, we analyze the dropzone-target inter-relationship and their affinity.
We observe that the list of targets is shared between attackers or compiled using similar conditions on IoT search engines like Shodan or Censys. For our geographical analysis, we comprehensively analyzed the distribution of the number of dropzones and targets (country, state, and city level). We visualize the target areas representing dropzone locations and their size scaled by the number of associated targets.

We utilized IoT search engines to conduct network penetration analysis of the dropzones and target IPs. We extract information such as the organization, the number of active ports, and vulnerabilities associated with the IP addresses; knowing which network ports are open on an IP address allows attackers to exploit them for DDoS attacks. Our analysis of the vulnerabilities associated with dropzone IP addresses, for example, reveals the level of risk involved and which malware instances they are associated with. Organizations with a high number of dropzone addresses indicate a tolerance for malicious activities and a susceptibility to compromise.

Our distributed analysis shows the exposure of endpoints correlating to the risk they possess. These endpoints need to be individually analyzed to extract patterns for predicting the chances of them being compromised. This issue will be our future work along with de-obfuscating functions to understand dynamic IP generation by malware.


\balance
\bibliographystyle{ACM-Reference-Format}
\bibliography{ref,conf}

\end{document}